\documentclass[a4paper,12pt]{article}

\usepackage{graphics}
\usepackage{latexsym}
\usepackage{amsmath}
\usepackage{amssymb}
\newcommand{\cA}{\mathcal{A}}

\newcommand{\cC}{\mathcal{C}}
\newcommand{\cD}{\mathcal{D}}

\newcommand{\cF}{\mathcal{F}}

\newcommand{\cI}{\mathcal{I}}

\newcommand{\cN}{\mathcal{N}}
\newcommand{\cO}{\mathcal{O}}

\newcommand{\cQ}{\mathcal{Q}}
\newcommand{\cR}{\mathcal{R}}

\newcommand{\zetto}{\mathbb{Z}}
\newcommand{\aaru}{\mathbb{R}}

\newcommand{\gh}{\#_{\mathrm{gh}}}

\newcommand{\llk}{\langle\!\langle}
\newcommand{\rrk}{\rangle\!\rangle}
\newcommand{\bllk}{\biggl\langle\!\!\!\biggl\langle}
\newcommand{\brrk}{\biggr\rangle\!\!\!\biggr\rangle}

\newcommand{\Qmid}{\cQ_{\mathrm{GRSZ}}}
\newcommand{\Qe}{Q_{\mathrm{even}}}
\newcommand{\Qo}{Q_{\mathrm{odd}}}
\newcommand{\Qev}{\cQ_{\mathrm{even}}}
\newcommand{\Qod}{\cQ_{\mathrm{odd}}}

\newcommand{\Qabc}{Q_{\mathrm{ABC}}}
\newcommand{\SIG}{\Sigma_{\epsilon}}
\newcommand{\sectiono}[1]{\section{#1}\setcounter{equation}{0}}

\begin{document}
\begin{titlepage}
\thispagestyle{empty}
\begin{flushright}
UT-02-42 \\
hep-th/0208009 \\
\end{flushright}

\vskip 1.5 cm

\begin{center}
\noindent{\textbf{\LARGE{On Ghost Structure of 
\vspace{0.5cm}\\ Vacuum Superstring Field Theory }}} 
\vskip 1.5cm
\noindent{\large{Kazuki Ohmori}\footnote{E-mail: 
ohmori@hep-th.phys.s.u-tokyo.ac.jp}}\\ 
\vspace{1cm}
\noindent{\small{\textit{Department of Physics, Faculty of Science, University of 
Tokyo}} \\ \vspace{2mm}
\small{\textit{Hongo 7-3-1, Bunkyo-ku, Tokyo 113-0033, Japan}}}
\end{center}
\vspace{1cm}
\begin{abstract}
After discussing the general form of the kinetic operator $\widehat{Q}$ around the tachyon vacuum, 
we determine the specific form of the pure-ghost kinetic operator $\widehat{\cQ}$ by requiring 
the twist invariance of the action. We obtain a novel result that the Grassmann-even piece $\Qev$ of $\widehat{\cQ}$ 
must act differently on GSO($+$) sector and on GSO($-$) sector to preserve the twist invariance, and show that  
this structure is crucial for gauge invariance of the action. With this choice of $\widehat{\cQ}$, 
we construct a solution in an approximation scheme which is conjectured to correspond to a non-BPS D9-brane. 
We consider both 0-picture cubic and Berkovits' non-polynomial superstring field theories for the NS sector. 
\end{abstract}
\end{titlepage}
\newpage
\baselineskip 6mm


\sectiono{Introduction}\label{sec:introduction}
For the past one and a half years vacuum string field theory (VSFT)~\cite{RSZ1,RSZ5} has been studied intensively. 
The distinguishing feature of this theory is that the kinetic operator is made purely out of ghost operators 
so that its equation of motion allows us to have solutions of matter-ghost factorized form. Since in bosonic 
string theory there is essentially only one kind of D-brane, it has been assumed that D-brane solutions in VSFT 
have the factorized form $\Psi=\Psi_{\mathrm{m}}\otimes\Psi_{\mathrm{g}}$ and that its ghost part $\Psi_{\mathrm{g}}$ 
is common to all D$p$-branes~\cite{RSZ2}. This assumption has successfully been tested by showing that such solutions 
correctly reproduce the ratios of D-brane tensions, in which only the matter part of the solutions was needed. 
As a matter of fact, many encouraging results have been obtained from the analysis of the matter sector: 
not only the ratios of tensions~\cite{RSZ2,RSZ4,Mukho,Okuyama3,MRVY} and various D-brane configurations~\cite{RSZ3,RSZ4} 
but also the overall D25-brane tension~\cite{Okawa} and the fluctuation spectrum 
around the brane solution~\cite{RSZ4,HK,HM,RSZ6,RV1,RV2,Okawa}. On the other hand, it has not been discussed whether 
the solutions of the ghost equation of motion proposed in~\cite{HK,GRSZ1,Okuda,PR} are suited for the description 
of the D-brane or not. Anyway, in section~\ref{sec:solution} of the present paper we will give arguments based on 
the assumption that the twisted ghost sliver state introduced in~\cite{GRSZ1} really describes the 
universal ghost solution for bosonic D-branes. 

In the type II superstring case, there are two kinds of D-branes: stable BPS D-branes and unstable non-BPS ones. 
Because of the qualitative difference between these two families, even if all D-brane solutions in vacuum 
superstring field theory have the matter-ghost factorized form and each of the families has its universal 
ghost solutions, there is no apparent reason to assume that the two ghost solutions should also agree with 
each other. In case they are really different, we will be forced to consider the full matter+ghost system 
even for the calculation of ratios of tensions between a BPS D-brane and a non-BPS D-brane. As another issue 
peculiar to the superstring case, there is a question of how the spacetime supersymmetry is restored around 
the tachyon vacuum: After the non-supersymmetric non-BPS D-branes or brane--antibrane systems have completely 
decayed through the tachyon condensation, it is believed that there remains the true vacuum for the type II 
closed superstrings without any D-branes, where $d=10,\cN=2$ spacetime supersymmetry should exist. 
Considering that the exact tachyon vacuum solution has not been obtained so far and that it seems difficult 
to investigate the supersymmetric structure in the level truncation scheme, the only way to proceed is to 
construct superstring field theory around the tachyon vacuum directly. 
Although we have no idea up to now how this phenomenon of supersymmetry restoration 
should be described in terms of \textit{open} string 
degrees of freedom, if we are to take the vacuum string field theory proposal seriously this is a problem of 
the kind that should be resolved within this theory. For this purpose, it is obvious that we must reveal the 
complete structure of the theory including both Neveu-Schwarz (NS) and Ramond (R) sectors.\footnote{Spacetime 
supersymmetry transformations for open string fields containing both GSO($\pm$) sectors have been 
discussed by Yoneya~\cite{Yoneya} in the context of Witten's cubic open superstring field theory~\cite{Witten}.} 
As a first step 
toward this goal, we consider in this paper the NS sector only. The above two interesting problems have 
motivated us to study in detail the ghost structure of superstring field theory around the tachyon vacuum. 
\medskip

The study of ghost kinetic operator in vacuum superstring field theory was initiated by Aref'eva 
\textit{et. al.}~\cite{ABG,NSgs}. Those authors argued that the kinetic operator $\widehat{Q}$ around the 
tachyon vacuum may contain a Grassmann-even operator $\Qe$ that mixes the GSO($\pm$) sectors of the string 
field, and proposed two possible candidates for the pure-ghost kinetic operator $\widehat{\cQ}$ where $\Qev$ 
was represented by linear combinations of $\gamma(i)$ and $\gamma(-i)$. However, they have not shown 
on what principle they determined the precise form of $\widehat{\cQ}$. In the present paper we adopt as a 
guiding principle the `twist invariance' of the action, which exists in the original (bosonic, 0-picture 
cubic super and Berkovits' super) string field theories on D-branes, reinforced with the gauge invariance. 
In fact, if the tachyon vacuum solution is represented by a twist-even string field configuration as 
has been the case in the level truncation calculations, this twist symmetry should survive in  
vacuum string field theory action. As a result, 
we will put forward a candidate different from theirs, and show that our choice gives rise to 
consistent gauge invariant actions. 
\medskip

This paper is organized as follows. In section~\ref{sec:proposal} we fix the form of the ghost kinetic operator 
$\widehat{\cQ}$ from the requirement of the twist invariance of the cubic action, and then examine its properties. 
In section~\ref{sec:solution} we try to solve the equations of motion, but we find a solution only in an 
approximation scheme. In section~\ref{sec:nonpoly} we extend the results obtained in the previous sections 
for the cubic theory to the non-polynomial superstring field theory by Berkovits. In section~\ref{sec:sumdis} 
we summarize our results and have discussion on further research. In Appendices we expose the technical details 
about the inner derivation formula for $\Qev$ and the twist invariance of our action.

\sectiono{Proposal for the (Super)ghost Kinetic Operator: Cubic Theory}\label{sec:proposal}
In sections \ref{sec:proposal} and \ref{sec:solution}, we consider the cubic vacuum superstring 
field theory~\cite{ABG,NSgs}. We first argue that a GSO($\pm$)-mixing Grassmann-even sector $\Qe$ 
generally arises in the kinetic operator $\widehat{Q}$ around the tachyon vacuum, 
following~\cite{ABG}, and that this is in fact necessary in order to have the expected structure of 
vacuum superstring field theory action. We then use the twist symmetry to determine the form of purely ghostly 
kinetic operator $\widehat{\cQ}$, and propose that this operator 
describes the open superstrings around one of the tachyon vacua. 

\subsection{Structure of the kinetic operator around the tachyon vacuum}\label{subsec:Qstr}
Let us begin with the cubic action for the Neveu-Schwarz sector string field $\widehat{A}$ 
in the background of a non-BPS D-brane~\cite{ABG}: 
\begin{equation}
S=-\frac{1}{2g_o^2}\mathrm{Tr}\left[\frac{1}{2}\llk \widehat{Y}_{-2}|\widehat{A},\widehat{Q}_B\widehat{A}
\rrk +\frac{1}{3}\llk\widehat{Y}_{-2}|\widehat{A},\widehat{A}*\widehat{A}\rrk\right], \label{eq:AA}
\end{equation}
where the \textit{internal Chan-Paton} structure is 
\begin{eqnarray}
\widehat{Q}_B&=&Q_B\otimes\sigma_3, \nonumber \\
\widehat{Y}_{-2}&=&Y(i)Y(-i)\otimes\sigma_3 \qquad\quad (Y(z)=c\partial\xi e^{-2\phi}(z)), \label{eq:AB} \\
\widehat{A}&=&A_+\otimes\sigma_3+A_-\otimes i\sigma_2, \nonumber 
\end{eqnarray}
and the trace Tr is taken over the space of these $2\times 2$ matrices ($\sigma_i$'s are the Pauli matrices). 
The GSO($+$) sector string field $A_+$ is Grassmann-odd and consists of states with integer weights, while 
the GSO($-$) field $A_-$ is Grassmann-even and has half-odd-integer weights. Both of them have ghost number 1 
and picture number 0. The bracket $\llk Y_{-2}|\ldots\rrk$ is defined in terms of the correlation functions 
in the world-sheet CFT as\footnote{Strictly speaking, the action of conformal transformations 
$f^{(n)}_k\circ A(0)=(f^{(n)\prime}_k(0))^hA(f^{(n)}_k(0))$ is not well-defined for vertex operators of 
non-integer weights $h$ unless we fix the phase ambiguity. We will mention it when it is needed.}
\begin{eqnarray}
\llk Y_{-2}|A_1,A_2\rrk &=& \left\langle Y(i)Y(-i) f^{(2)}_1\circ A_1(0)f^{(2)}_2\circ A_2(0)
\right\rangle_{\mathrm{UHP}}, \label{eq:AC} \\
\llk Y_{-2}|A_1,A_2*A_3\rrk &=& \left\langle Y(i)Y(-i) f^{(3)}_1\circ A_1(0)f^{(3)}_2\circ A_2(0)
f^{(3)}_3\circ A_3(0)\right\rangle_{\mathrm{UHP}}, \label{eq:AD} 
\end{eqnarray}
where 
\begin{eqnarray}
f^{(n)}_k(z)&=&h^{-1}\left(e^{2\pi i\frac{k-1}{n}}h(z)^{\frac{2}{n}}\right), \label{eq:AE} \\
& & \hspace{-2cm} h(z)=\frac{1+iz}{1-iz}, \qquad h^{-1}(z)=-i\frac{z-1}{z+1}, \nonumber 
\end{eqnarray}
and in particular, 
\begin{equation}
f^{(2)}_1(z)=z, \qquad f^{(2)}_2(z)=h^{-1}\left( e^{\pi i}h(z)\right)=-\frac{1}{z}\equiv I(z). \label{eq:AF}
\end{equation}
Note that since the inverse picture-changing operator $Y(z)$ is a primary field of conformal weight 0, 
it is not affected by conformal transformations that keep the open string midpoint $\pm i$ fixed. 
\medskip 

By taking the trace over the internal Chan-Paton matrices in advance, the action~(\ref{eq:AA}) can be 
rewritten explicitly in terms of $A_{\pm}$ as 
\begin{eqnarray}
S&=&-\frac{1}{g_o^2}\Biggl[\frac{1}{2}\llk Y_{-2}|A_+,Q_BA_+\rrk +\frac{1}{2}\llk Y_{-2}|A_-, Q_BA_-\rrk
\nonumber \\
& &\hspace{7mm}{}+\frac{1}{3}\llk Y_{-2}|A_+,A_+*A_+\rrk-\llk Y_{-2}|A_+,A_-*A_-\rrk\Biggr]. \label{eq:AG}
\end{eqnarray}
From this expression, one can immediately see that this action is invariant under the sign-flip of the 
GSO($-$) sector string field: $A_-\longrightarrow -A_-$. This symmetry guarantees that the effective potential 
for the `real'\footnote{Remember that in 0-picture cubic superstring field theory there also exists an 
`auxiliary' tachyon field in the GSO($+$) sector~\cite{0011117}.} tachyon field living in the GSO($-$) sector 
takes the left-right symmetric double-well form. 
Now, postulating the solution $\widehat{A}_0$ corresponding to one of the doubly-degenerate tachyon vacua, 
we expand the string field $\widehat{A}$ around it as $\widehat{A}=\widehat{A}_0+\widehat{a}$. 
Then part of the action quadratic in the fluctuation fields $\widehat{a}=a_+\otimes\sigma_3+a_-\otimes i\sigma_2$ 
is found to be 
\begin{eqnarray}
-g_o^2S_{\mathrm{quad}}&=&\frac{1}{2}\llk Y_{-2}|a_+,(Q_Ba_++A_{0+}*a_++a_+*A_{0+}-A_{0-}*a_--a_-*A_{0-})\rrk
\label{eq:AH} \\ & &\hspace{-5mm}{}+\frac{1}{2}\llk Y_{-2}|a_-,(Q_Ba_-+A_{0+}*a_--a_-*A_{0+}-A_{0-}*a_++a_+*A_{0-})\rrk, 
\phantom{QQQ} \nonumber 
\end{eqnarray}
where we have used the cyclicity relation of the 3-string vertex: 
\begin{equation}
\llk Y_{-2}|A_1,A_2*A_3\rrk=e^{2\pi ih_{A_3}}\llk Y_{-2}|A_3,A_1*A_2\rrk \label{eq:AI}
\end{equation}
with $h_{A_3}$ denoting the conformal weight of $A_3$. That is to say, if the vertex operator $A_3$ to be moved 
is in the GSO($-$) sector an additional minus sign arises.\footnote{I'd like to thank I. Kishimoto for reminding 
me of this fact.} This is due to the $2\pi$-rotation $\cR_{2\pi}(z)\equiv h^{-1}(e^{2\pi i}h(z))$ of the unit 
disk, which provides a conformal factor $[\cR^{\prime}_{2\pi}(0)]^{h_{A_3}}=e^{2\pi ih_{A_3}}$~\cite{BSZ,0011117}. 
Note that~(\ref{eq:AH}) contains cross-terms among $a_+$ and $a_-$ through the help of $A_{0-}$. 
In the matrix notation~(\ref{eq:AA}), the same quadratic action is also written as 
\begin{equation}
-g_o^2S_{\mathrm{quad}}=\frac{1}{4}\mathrm{Tr}\llk\widehat{Y}_{-2}|\widehat{a}, 
(\widehat{Q}_B\widehat{a}+\widehat{A}_0*\widehat{a}+\widehat{a}*\widehat{A}_0)\rrk, \label{eq:AJ}
\end{equation}
where we have used the fact that the hatted fields satisfy the cyclicity relation without any sign factor 
because the $-$ sign for the GSO($-$) field appearing in~(\ref{eq:AI}) is compensated for by one more sign arising from 
$i\sigma_2\widehat{Y}_{-2}=-\widehat{Y}_{-2}i\sigma_2$. 
Here we define the new kinetic operator around the tachyon vacuum as 
\begin{equation}
\widehat{Q}\widehat{\Phi}\equiv \widehat{Q}_B\widehat{\Phi}+\widehat{A}_0*\widehat{\Phi}-(-1)^{\gh(\widehat{\Phi})}
\widehat{\Phi}*\widehat{A}_0, \label{eq:AK}
\end{equation}
where $\gh(\widehat{\Phi})$ denotes the ghost number of $\widehat{\Phi}$. 
To understand why we have adopted the ghost number grading instead of the Grassmannality, recall that 
the internal Chan-Paton factors have originally been introduced in such a way that the GSO($\pm$) string 
fields with different Grassmannalities obey the same algebraic relations~\cite{BSZ}. 
Note that if we restrict ourselves to the GSO($+$) states, these two gradings agree with each other: 
$(-1)^{\gh(\Phi_+)}=(-1)^{|\Phi_+|}$.\footnote{Generically a relation 
$(-1)^{\gh(\widehat{\Phi})}(-1)^{|\widehat{\Phi}|}(-1)^{\mathrm{GSO}(\widehat{\Phi})}=1$ holds among 
the ghost number $\gh(\widehat{\Phi})$, Grassmannality $|\widehat{\Phi}|$ and GSO parity 
$\mathrm{GSO}(\widehat{\Phi})$ of $\widehat{\Phi}$.}
By using $\widehat{Q}$ defined in~(\ref{eq:AK}), eq.(\ref{eq:AJ}) can further be rewritten as 
\begin{equation}
-g_o^2S_{\mathrm{quad}}=\frac{1}{2}\llk Y_{-2}|a_+,\frac{1}{2}\mathrm{Tr}(\widehat{Q}\widehat{a})\rrk +
\frac{1}{2}\llk Y_{-2}|a_-,\frac{1}{2}\mathrm{Tr}(\sigma_1\widehat{Q}\widehat{a})\rrk. \label{eq:AL}
\end{equation}
Comparing eqs.(\ref{eq:AH}) and (\ref{eq:AL}), we notice that $\mathrm{Tr}(\widehat{Q}i\sigma_2)a_-$ and 
$\mathrm{Tr}(\sigma_1\widehat{Q}\sigma_3)a_+$ (as well as $\mathrm{Tr}(\widehat{Q}\sigma_3)a_+$ and 
$\mathrm{Tr}(\sigma_1\widehat{Q}i\sigma_2)a_-$) must be non-zero in general, because the tachyon vacuum solution 
$\widehat{A}_0$ contains non-zero GSO($-$) components $A_{0-}$ and there is no reason that 
$-A_{0-}*a_--a_-*A_{0-}$ and $-A_{0-}*a_++a_+*A_{0-}$ should vanish in~(\ref{eq:AH}). 
This can be achieved by letting $\widehat{Q}$ have the following internal Chan-Paton structure~\cite{ABG}: 
\begin{equation}
\widehat{Q}=\Qo\otimes\sigma_3-\Qe\otimes i\sigma_2, \label{eq:AM}
\end{equation}
where $\Qo$ and $\Qe$ are Grassmann-odd and Grassmann-even operators respectively. 
Explicit actions of $\Qo,\Qe$ can be found by comparing the both sides of eq.(\ref{eq:AK}). When $\widehat{\cQ}$ 
acts on a string field $\widehat{X}$ of \textit{odd ghost number} of the form 
\begin{equation}
\widehat{X}=X_+\otimes\sigma_3+X_-\otimes i\sigma_2, \label{eq:Xodd}
\end{equation}
we find 
\begin{eqnarray}
\Qo X_+&=&Q_BX_++A_{0+}*X_++X_+*A_{0+}, \nonumber \\
\Qo X_-&=&Q_BX_-+A_{0+}*X_--X_-*A_{0+}, \label{eq:QXodd} \\
\Qe X_+&=&-A_{0-}*X_++X_+*A_{0-}, \nonumber \\
\Qe X_-&=&-A_{0-}*X_--X_-*A_{0-}, \nonumber 
\end{eqnarray}
where the GSO($+$) component $X_+$ is Grassmann-odd and the GSO($-$) component $X_-$ is Grassmann-even. 
On the other hand, when $\widehat{\cQ}$ acts on a string field $\widehat{Y}$ of \textit{even ghost number} 
having the form
\begin{equation}
\widehat{Y}=Y_+\otimes\mathbf{1}+Y_-\otimes\sigma_1, \label{eq:Yeven}
\end{equation}
we obtain 
\begin{eqnarray}
\Qo Y_+&=&Q_BY_++A_{0+}*Y_+-Y_+*A_{0+}, \nonumber \\
\Qo Y_-&=&Q_BY_-+A_{0+}*Y_-+Y_-*A_{0+}, \label{eq:QYeven} \\
\Qe Y_+&=&-A_{0-}*Y_++Y_+*A_{0-}, \nonumber \\
\Qe Y_-&=&-A_{0-}*Y_--Y_-*A_{0-}, \nonumber 
\end{eqnarray}
where $Y_+$ is Grassmann-even and $Y_-$ is Grassmann-odd. 
These two sets (\ref{eq:QXodd}), (\ref{eq:QYeven}) of equations can be written in a unified manner as 
\begin{eqnarray}
& &\Qo a=Q_Ba+A_{0+}*a-(-1)^{|a|}a*A_{0+}, \label{eq:AN} \\
& &\Qe a=-A_{0-}*a+(-1)^{\mathrm{GSO}(a)}a*A_{0-}, \label{eq:AO}
\end{eqnarray}
for $a$ of any ghost number and of any GSO parity, 
where $|a|$ denotes the Grassmannality of $a$ ($|a|=0/1$ mod 2 if $a$ is Grassmann-even/odd) and 
$\mathrm{GSO}(a)$ represents the GSO parity of $a$ ($(-1)^{\mathrm{GSO}(a)}=\pm 1$ if $a$ is in the GSO($\pm$) sector). 
The difference in the Chan-Paton structures for even and odd ghost number string fields, (\ref{eq:Xodd}) and 
(\ref{eq:Yeven}),\footnote{The same assignment of the Chan-Paton matrices to string fields 
had been proposed in~\cite{Smet}.} 
comes from the consistency of the $*$-product: For example, $*$-multiplication of two string fields 
both having ghost number 1 must give rise to a structure appropriate for a string field of ghost number 2, and 
the set of all ghost number 0 string fields must form a closed subalgebra under the $*$-product. 
\medskip

\begin{figure}[htbp]
  \begin{center}
    \scalebox{0.8}[0.7]{\includegraphics{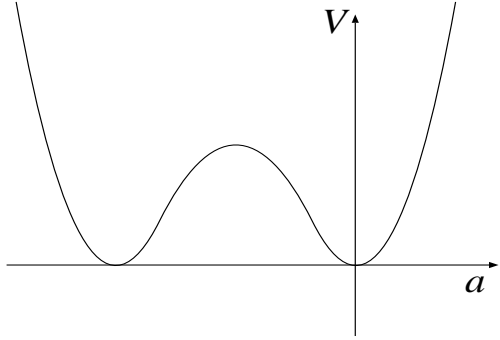}}
  \end{center}
  \caption{A schematic picture of the tachyon potential. No reflection symmetry is expected around 
  the tachyon vacuum ($a=0$).}
  \label{fig:pot}
\end{figure}

Here we argue that non-zero $\Qe$ is necessary to have a sensible vacuum superstring field theory. 
To this end, suppose that we are given $\Qe =0$, so $\widehat{Q}=\Qo \otimes\sigma_3$. 
Since this $\widehat{Q}$ has the same structure as $\widehat{Q}_B=Q_B\otimes\sigma_3$, the action 
expanded around the tachyon vacuum would again take the same form as the original one~(\ref{eq:AG}) 
with $Q_B$ replaced by $\Qo$ and $A_{\pm}$ denoting the fluctuations around the tachyon vacuum. 
Then, for the same reason as the one mentioned above, the `string field theory around the tachyon 
vacuum' would have a $\zetto_2$-reflection symmetry under $a_-\longrightarrow -a_-$, which means that 
if we have a solution $a_+\otimes\sigma_3+a_-\otimes i\sigma_2$ then we find one more solution 
$a_+\otimes\sigma_3-a_-\otimes i\sigma_2$ with the same energy density. However, we do not expect 
such a degeneracy of solutions to exist in vacuum superstring field theory (Figure~\ref{fig:pot}).
There remains a possibility that any relevant solutions in this theory, such as D-branes, consist 
only of GSO($+$) components so that we can avoid having a pair of degenerate solutions, but 
we do not believe that this is the case. Next we show that $\Qe$ plays the r\^{o}le of removing this 
unwanted degeneracy. Making use of the new kinetic operator $\widehat{Q}$ defined in~(\ref{eq:AK}) or 
(\ref{eq:AN})--(\ref{eq:AO}), we can write the cubic superstring field theory action around the 
tachyon vacuum as 
\begin{eqnarray}
S&=&-\frac{1}{g_o^2}\Biggl[\frac{1}{2}\llk Y_{-2}|a_+,\Qo a_+\rrk+\frac{1}{2}\llk Y_{-2}|a_-,\Qo a_-\rrk
\nonumber \\ & &\hspace{7mm}{}+\frac{1}{2}\llk Y_{-2}|a_+,\Qe a_-\rrk+\frac{1}{2}\llk Y_{-2}|
a_-,\Qe a_+\rrk \label{eq:AP} \\ & &\hspace{7mm}{}+\frac{1}{3}\llk Y_{-2}|a_+,a_+*a_+\rrk -\llk Y_{-2}|
a_+,a_-*a_-\rrk \Biggr] \nonumber \\
&=&-\frac{1}{2g_o^2}\mathrm{Tr}\left[\frac{1}{2}\llk \widehat{Y}_{-2}|\widehat{a},\widehat{Q}\widehat{a}
\rrk+\frac{1}{3}\llk\widehat{Y}_{-2}|\widehat{a},\widehat{a}*\widehat{a}\rrk\right], \label{eq:AQ}
\end{eqnarray}
where we have omitted a constant term. In the second line of eq.(\ref{eq:AP}) the GSO($-$) string field 
$a_-$ enters the action linearly, so that the above-mentioned $\zetto_2$ symmetry is absent in this 
action.\footnote{This argument relies on the very fundamental assumption that in vacuum superstring 
field theory we do not perform the GSO-projection on the open string field, 
as in the case of the open superstring theory on 
non-BPS D-branes.} At this stage the kinetic operator $\widehat{Q}$ is regular and is not considered to be 
pure ghost. In the next subsection we will try to determine the form of the purely ghostly 
kinetic operator $\widehat{\cQ}$ which is supposed to arise after a suitable singular field redefinition. 

\subsection{Ghost kinetic operator of cubic vacuum superstring field theory}\label{subsec:Qform}
Let us briefly review the argument of Gaiotto, Rastelli, Sen and Zwiebach~\cite{GRSZ1} about the 
origin of a pure-ghost kinetic operator of bosonic vacuum string field theory. 
First assume that a regular representative $Q$ of an equivalence class of kinetic operators around the 
tachyon vacuum takes the following form 
\begin{equation}
Q=\int_{-\pi}^{\pi}d\sigma\ a_c(\sigma)c(\sigma)+\sum_r\int_{-\pi}^{\pi}d\sigma\ a_r(\sigma)
O_r(\sigma), \label{eq:AR}
\end{equation}
where $a_{c,r}$ are functions of $\sigma$ and $O_r$'s are local operators of ghost number 1 with 
conformal weights higher than that of $c$. Then consider performing a reparametrization of the open 
string coordinate: $\sigma\to f(\sigma)$, which keeps the open string midpoint $\pm\pi /2$ fixed and 
is symmetric about it. While this operation does not change the $*$-product, it induces a transformation 
on the operator~(\ref{eq:AR}) as 
\begin{equation}
Q\longrightarrow\cQ =\int_{-\pi}^{\pi}d\sigma\ a_c(\sigma)(f^{\prime}(\sigma))^{-1}c(f(\sigma))+\sum_r
\int_{-\pi}^{\pi}d\sigma\ a_r(\sigma)(f^{\prime}(\sigma))^{h_r}O_r(f(\sigma)). \label{eq:AS}
\end{equation}
If we choose $f(\sigma)$ such that $f^{\prime}(\sigma)\simeq (\sigma\mp\frac{\pi}{2})^2+\varepsilon_r^2$ 
near $\sigma=\pm\pi /2$ with small $\varepsilon_r$, the integrand of the first term becomes large around 
$\sigma=\pm\pi /2$ and, in the limit $\varepsilon_r\to 0$, all other contributions can be neglected. In this way  
we have obtained simple but singular constituents of pure-ghost kinetic operator: $\varepsilon_r^{-1}c(i)$ 
and $\varepsilon_r^{-1}c(-i)$ in the upper half plane coordinate. The relative 
coefficient between these two terms will be fixed by requiring that the kinetic operator $\cQ$ preserve the 
twist invariance of the action.

It is known that the bosonic cubic open string field theory action 
\begin{equation}
S(\Phi)=-\frac{1}{g_o^2}\left[\frac{1}{2}\langle\Phi,Q_B\Phi\rangle+\frac{1}{3}\langle
\Phi,\Phi *\Phi\rangle\right] \label{eq:AT}
\end{equation}
has a twist symmetry~\cite{GZ,Zwiebach}. On an arbitrary $L_0^{\mathrm{tot}}$-eigenstate $|\Phi\rangle$ 
the twist operator $\Omega$ acts as 
\begin{equation}
\Omega |\Phi\rangle =(-1)^{h_{\Phi}+1}|\Phi\rangle, \label{eq:AU}
\end{equation}
where $h_{\Phi}$ is the $L_0^{\mathrm{tot}}$-eigenvalue of $|\Phi\rangle$. It can be shown that the action~(\ref{eq:AT}) 
is twist-invariant, $S(\Omega\Phi)=S(\Phi)$~\cite{Zwiebach}. Thanks to this property, we could restrict the string 
field to be twist-even in computing the tachyon potential~\cite{potential}. The proof of the twist-invariance of the 
action uses the fact that the usual BRST operator $Q_B$ commutes with the twist operator $\Omega$: 
\begin{equation}
\Omega (Q_B|\Phi\rangle)=Q_B(\Omega |\Phi\rangle). \label{eq:AV}
\end{equation}
This property holds because $Q_B$ is the zero-mode of the BRST current $j_B$ so that it does not change 
the $L_0^{\mathrm{tot}}$-eigenvalue of the state. Let us now turn to bosonic vacuum string field theory 
whose action is given by 
\begin{equation}
S_V(\Psi)=-\kappa_0\left[\frac{1}{2}\langle\Psi,\cQ\Psi\rangle+\frac{1}{3}\langle\Psi,\Psi *\Psi\rangle
\right], \label{eq:AW}
\end{equation}
where, according to the arguments of the last paragraph, the kinetic operator $\cQ$ is some linear 
combination of $c(i)$ and $c(-i)$. Since the original action~(\ref{eq:AT}) has the twist symmetry and 
the tachyon vacuum solution is believed to be represented by a twist-even configuration~\cite{potential},  
it is natural to assume that the VSFT action~(\ref{eq:AW}) also has twist symmetry. 
For this action to be twist invariant, $\cQ$ must commute with $\Omega$: 
\begin{equation}
\Omega (\cQ |\Psi\rangle)=\cQ (\Omega |\Psi\rangle). \label{eq:AX}
\end{equation}
Since we have 
\[ \Omega (c_n|\Psi\rangle)=(-1)^{(h_{\Psi}-n)+1}(c_n|\Psi\rangle)=(-1)^{-n}c_n(\Omega|\Psi\rangle), \]
$\cQ$ satisfies the twist-invariance condition~(\ref{eq:AX}) if $\cQ$ consists of 
even modes $c_{2n}$ only. This requirement uniquely fixes the relative normalization as~\cite{GRSZ1} 
\begin{eqnarray}
\cQ=\Qmid &\equiv& \frac{1}{2i}(c(i)-c(-i)) \label{eq:AY} \\
&=&c_0-(c_2+c_{-2})+(c_4+c_{-4})-\ldots, \nonumber
\end{eqnarray}
where an overall normalization constant has been absorbed into the definition of the string field. 
This kinetic operator was shown~\cite{GRSZ1,Okuyama2} to agree with the one found in~\cite{HK} by requiring that 
the Siegel gauge solution should solve the equations of motion in the full gauge-unfixed field 
configuration space. 
\medskip

\begin{figure}[htbp]
	\begin{center}
	\scalebox{0.7}[0.7]{\includegraphics{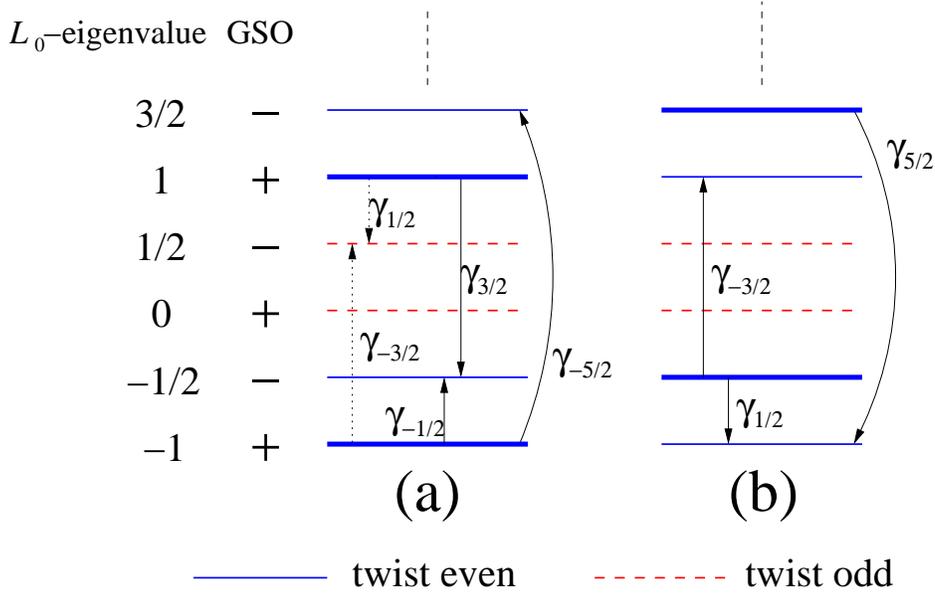}}
	\end{center}
	\caption{(a) The action of $\gamma_r$ on GSO($+$) twist-even states. $\gamma_r$ with $r\in 2\zetto-\frac{1}{2}$ 
	preserve the twist eigenvalues (indicated by solid arrows), whereas the wrong ones with 
	$r\in 2\zetto +\frac{1}{2}$ reverse the twist eigenvalues (dotted arrows). (b) The action of $\gamma_r$ on 
	GSO($-$) twist-even states. Now it is $\gamma_r$ with $r\in 2\zetto +\frac{1}{2}$ that preserve the twist.}
	\label{fig:level}
\end{figure}

Now we turn to the superstring case, where there are two negative-dimensional operators $c$ and $\gamma$. 
Suppose that after a reparametrization of the string coordinate implemented by a function $f$, 
the kinetic operator is written as 
\begin{eqnarray}
\widehat{\cQ}&=&\int_{-\pi}^{\pi}d\sigma\ a_c(\sigma)[f^{\prime}(\sigma)]^{-1}c(f(\sigma))\otimes \sigma_3 
\nonumber \\ & &+\int_{-\pi}^{\pi}d\sigma\ a_{\gamma}(\sigma)[f^{\prime}(\sigma)]^{-\frac{1}{2}}\gamma
(f(\sigma))\otimes i\sigma_2+\ldots. \label{eq:IC}
\end{eqnarray}
Let us postulate a function $f$ which around $\sigma=\frac{\pi}{2}$ behaves as 
\[ [f^{\prime}(\sigma)]^{-\frac{1}{2}}\sim \frac{1}{\varepsilon_r}\delta\left(\sigma -\frac{\pi}{2}
\right) \qquad \mbox{ and } \qquad  [f^{\prime}(\sigma)]^{-1}\sim \frac{1}{\varepsilon_r^2}
\delta\left(\sigma -\frac{\pi}{2}\right) \]
in the singular limit $\varepsilon_r\to 0$. 
If we take such $f$ that behaves similarly near $\sigma=-\frac{\pi}{2}$ and is regular everywhere except 
at $\sigma=\pm\frac{\pi}{2}$, then $\widehat{\cQ}$~(\ref{eq:IC}) in the limit $\varepsilon_r\to 0$ 
is dominated by 
\begin{eqnarray}
\widehat{\cQ}&=&\frac{1}{\varepsilon_r^2}\left(a_c\left(\frac{\pi}{2}\right)c\left(\frac{\pi}{2}\right)
+a_c\left(-\frac{\pi}{2}\right)c\left(-\frac{\pi}{2}\right)\right)\otimes\sigma_3 \label{eq:IF} \\
& &+\frac{1}{\varepsilon_r}\left(a_{\gamma}\left(\frac{\pi}{2}\right)\gamma\left(\frac{\pi}{2}\right)
+a_{\gamma}\left(-\frac{\pi}{2}\right)\gamma\left(-\frac{\pi}{2}\right)\right)\otimes i\sigma_2, \nonumber 
\end{eqnarray}
where we have used $f\left(\pm\frac{\pi}{2}\right)=\pm\frac{\pi}{2}$. We then require $\widehat{\cQ}$ to 
preserve the twist invariance of the action, by which the form of $\widehat{\cQ}$ can further be restricted 
without knowing the precise values of $a_{c,\gamma}\left(\pm\frac{\pi}{2}\right)$. 
In 0-picture cubic superstring field theory, it has been shown in~\cite{0011117} that 
the action~(\ref{eq:AA}) is invariant under the twist operation 
\begin{equation}
\Omega |A\rangle =\left\{
  \begin{array}{ll}
     (-1)^{h_A+1}|A\rangle  &  \mbox{for GSO($+$) states } (h_A\in \zetto)  \\
     (-1)^{h_A+\frac{1}{2}}|A\rangle  &  \mbox{for GSO($-$) states } (h_A\in\zetto+\frac{1}{2}).  
  \end{array}
\right. \label{eq:AZ}
\end{equation}
Since $\Qmid$~(\ref{eq:AY}) preserves the twist eigenvalues on both GSO($\pm$) sectors and 
is Grassmann-odd, the odd part $\Qo$ of the kinetic operator~(\ref{eq:AM}), 
after the singular reparametrization, becomes 
\begin{equation}
\Qo\longrightarrow \Qod =\frac{1}{2i\varepsilon_r^2}(c(i)-c(-i)) \qquad\quad (\varepsilon_r\to 0), \label{eq:BA}
\end{equation}
where we have made a finite rescaling of $\varepsilon_r$ for convenience. 
On the other hand, since $\gamma(z)$ 
has half-odd-integer modes in the NS sector and mixes the GSO($\pm$) sectors, its twist property 
becomes much more complicated. For example, let us consider a GSO($+$) state $|A_+\rangle$ with 
$L_0^{\mathrm{tot}}$-eigenvalue $h_+$. From eq.(\ref{eq:AZ}), we have 
\begin{equation}
\Omega |A_+\rangle=(-1)^{h_++1}|A_+\rangle. \label{eq:BB1}
\end{equation}
When $\gamma_r$ $(r\in\zetto +\frac{1}{2})$ acts on $|A_+\rangle$, the resulting state $\gamma |A_+\rangle$ 
is in the GSO($-$) sector and hence its twist eigenvalue must be evaluated as a GSO($-$) state: 
\begin{equation}
\Omega (\gamma_r|A_+\rangle)=(-1)^{(h_+-r)+\frac{1}{2}}(\gamma_r|A_+\rangle). \label{eq:BB}
\end{equation}
Combining eqs.(\ref{eq:BB1}) and (\ref{eq:BB}), we find the following relation: 
\begin{equation}
\Omega (\gamma_r|A_+\rangle)=(-1)^{-r-\frac{1}{2}}\gamma_r(\Omega|A_+\rangle). \label{eq:BB2}
\end{equation}
Thus we conclude that $\gamma_r$ acting on a GSO($+$) state $|A_+\rangle$ commutes with the twist 
\begin{equation}
\Omega (\gamma_r|A_+\rangle)=\gamma_r(\Omega|A_+\rangle) \label{eq:BC}
\end{equation}
\textit{when} $\mathit{r\in 2\zetto-\frac{1}{2}}$. This argument is visualized in Figure~\ref{fig:level}(a). 
Since we find 
\begin{equation}
\frac{1}{2i}(\gamma(i)-\gamma(-i))=\frac{1}{2i}\left(\sum_{r\in\zetto +\frac{1}{2}}\frac{\gamma_r}{i^{r-\frac{1}{2}}}
-\sum_{r\in\zetto +\frac{1}{2}}\frac{\gamma_r}{(-i)^{r-\frac{1}{2}}}\right)=\sum_{n\in\zetto}(-1)^n
\gamma_{-\frac{1}{2}+2n}, \label{eq:BD}
\end{equation}
we identify a candidate for the twist-preserving kinetic operator as 
\begin{equation}
\Qe\longrightarrow\cQ_{\mathrm{even}}^{\mathrm{GSO}(+)}=\frac{q_1}{2i\varepsilon_r}(\gamma(i)-\gamma(-i)) 
\qquad\quad (\varepsilon_r\to 0), \label{eq:BE}
\end{equation}
where $q_1$ is a finite real constant. However, it turns out that this kinetic operator, when acting on a 
GSO($-$) state, does \textit{not} preserve the twist eigenvalue. To see this, consider a GSO($-$) state 
$|A_-\rangle$ with $L_0^{\mathrm{tot}}$-eigenvalue $h_-$. From the relation 
\begin{equation}
\Omega (\gamma_r|A_-\rangle)=(-1)^{(h_--r)+1}(\gamma_r|A_-\rangle)=(-1)^{-r+\frac{1}{2}}\gamma_r
(\Omega |A_-\rangle), \label{eq:BF}
\end{equation}
$\gamma_r$ commutes with $\Omega$ if $r\in 2\zetto +\frac{1}{2}$, rather than $r\in 2\zetto -\frac{1}{2}$ 
(see Figure~\ref{fig:level}(b)). Therefore, the twist-preserving kinetic operator acting on a GSO($-$) 
state should take the form 
\begin{eqnarray}
\Qe\longrightarrow \cQ_{\mathrm{even}}^{\mathrm{GSO}(-)}&=&\frac{q_2}{2\varepsilon_r}(\gamma(i)+\gamma(-i))
\label{eq:BG} \\ &=&\frac{q_2}{\varepsilon_r}\sum_{n\in\zetto}(-1)^n\gamma_{\frac{1}{2}+2n} \nonumber
\end{eqnarray}
in the $\varepsilon_r\to 0$ limit, where $q_2$ is another constant. 

Our proposal that the kinetic operator $\Qev$ takes different forms~(\ref{eq:BE}), (\ref{eq:BG}) depending 
on the GSO parity of the states on which $\Qev$ acts may seem strange, but such a behavior is in fact necessary for 
the construction of gauge-invariant actions: 
To show the gauge invariance of the action, we need the hermiticity relation for 
$\widehat{\cQ}=\Qod\otimes\sigma_3-\Qev\otimes i\sigma_2$,\footnote{Since 
eq.(\ref{eq:BK}) is derived from the derivation property~(\ref{eq:CL}) of $\widehat{\cQ}$ and the fact that 
$\widehat{\cQ}$ annihilates $\langle\cI|Y_{-2}$, self-consistency also requires eq.(\ref{eq:BK}) to hold.} 
\begin{equation}
\llk\widehat{Y}_{-2}|\widehat{\cQ}\widehat{A},\widehat{B}\rrk =-(-1)^{\gh (\widehat{A})}\llk \widehat{Y}_{-2}|
\widehat{A},\widehat{\cQ}\widehat{B}\rrk. \label{eq:BK}
\end{equation}
Given the internal Chan-Paton structure 
\begin{eqnarray}
\widehat{A}=A_+\otimes\sigma_3+A_-\otimes i\sigma_2 &\quad& \mbox{for } \gh (\widehat{A}) \mbox{ odd} \nonumber \\
\widehat{A}=A_+\otimes\mathbf{1}+A_-\otimes\sigma_1 &\quad& \mbox{for } \gh (\widehat{A}) \mbox{ even}, \label{eq:BL}
\end{eqnarray}
(similarly for $\widehat{B}$)
and concentrating on the $\Qev$ part, eq.(\ref{eq:BK}) is rewritten as 
\begin{eqnarray}
& &\llk Y_{-2}|\Qev A_+,B_-\rrk=-\llk Y_{-2}|A_+,\Qev B_-\rrk, \label{eq:BM2} \\
& &\llk Y_{-2}|\Qev A_-,B_+\rrk= \llk Y_{-2}|A_-,\Qev B_+\rrk. \label{eq:BM3}
\end{eqnarray}
Here, let us closely look at the 2-point vertex (BPZ inner product)  
\[ \llk Y_{-2}|A_1,A_2\rrk=\langle Y(i)Y(-i) A_1(0)I\circ A_2(0)\rangle_{\mathrm{UHP}}. \]
The inversion $I(z)=-\frac{1}{z}=h^{-1}(-h(z))$ can be written in the following two ways: \linebreak
$\cR_{\pi}(z)\equiv h^{-1}(e^{\pi i}h(z))$ and $\cR_{-\pi}(z)\equiv h^{-1}(e^{-\pi i}h(z))$. Recalling that the 
$2\pi$-rotation $\cR_{2\pi}(z)=h^{-1}(e^{2\pi i}h(z))\simeq z$ acts non-trivially on GSO($-$) states with 
half-integer weights, we must define the action of the inversion $I(z)=-1/z$ on $A_2$ more precisely. 
We adopt the convention 
\begin{eqnarray}
\llk Y_{-2}|A_1,A_2\rrk&\equiv&\langle Y(i)Y(-i)\ A_1(0)\ \cR_{\pi}\circ A_2(0)\rangle_{\mathrm{UHP}} \label{eq:BI} \\
&=&[\cR_{\pi}^{\prime}(0)]^{h_{A_2}}\langle Y(i)Y(-i)\ A_1(0)\ A_2(\cR_{\pi}(0))\rangle_{\mathrm{UHP}}, \nonumber
\end{eqnarray}
where 
\begin{equation}
[\cR_{\pi}^{\prime}(z)]^h\equiv z^{-2h},\qquad [\cR_{-\pi}^{\prime}(z)]^h\equiv e^{-2\pi ih}z^{-2h}
=(-1)^{2h}z^{-2h}, \label{eq:BJ} 
\end{equation}
and we have assumed $A_2$ to be a primary field in (\ref{eq:BI}). The above definition~(\ref{eq:BJ}) is 
consistent with the composition laws 
\[ \cR_{\pi}\circ\cR_{\pi}(z)=\cR_{2\pi}(z), \qquad \cR_{\pi}\circ\cR_{-\pi}(z)=\cR_{-\pi}\circ\cR_{\pi}(z)=z, \]
and $[\cR_{2\pi}^{\prime}(z)]^h=e^{2\pi ih}$. Defined this way, the inner product~(\ref{eq:BI}) 
has been made well-defined for $\mathrm{GSO}(-)$ states as well. Now let us see eq.(\ref{eq:BM3}) in detail. 
The left-hand side of~(\ref{eq:BM3}) can be written as 
\begin{equation}
\llk Y_{-2}|\cQ_{\mathrm{even}}^{\mathrm{GSO}(-)}A_-,B_+\rrk=\langle Y(i)Y(-i)A_-(0)
\cQ_{\mathrm{even}}^{\mathrm{GSO}(-)}
(\cR_{\pi}\circ B_+(0))\rangle_{\mathrm{UHP}}, \label{eq:BN}
\end{equation}
while the right-hand side of~(\ref{eq:BM3}) is 
\begin{equation}
\llk Y_{-2}|A_-, \cQ_{\mathrm{even}}^{\mathrm{GSO}(+)}B_+\rrk=\llk Y(i)Y(-i)A_-(0)(\cR_{\pi}\circ 
\cQ_{\mathrm{even}}^{\mathrm{GSO}(+)})(\cR_{\pi}\circ B_+(0))\rangle_{\mathrm{UHP}}. \label{eq:BO}
\end{equation}
For these two expressions to agree with each other, we must have 
\begin{equation}
\cR_{\pi}\circ\cQ_{\mathrm{even}}^{\mathrm{GSO}(+)}=\cQ_{\mathrm{even}}^{\mathrm{GSO}(-)}, \label{eq:BP}
\end{equation}
but this equation \textit{cannot be satisfied if we stick to the case} 
$\cQ_{\mathrm{even}}^{\mathrm{GSO}(+)}=\cQ_{\mathrm{even}}^{\mathrm{GSO}(-)}$ because neither of 
$\gamma(\pm i), \gamma(i)\pm\gamma(-i)$ is self-conjugate under the inversion $\cR_{\pm\pi}$.\footnote{Generically, 
operators of half-integer weights satisfy $\cR_{\pi}\circ\cR_{\pi}\circ\cO=-\cO$ so that it seems 
impossible to construct  operators which are real and self-conjugate under $\cR_{\pi}$.} 
Thus we conclude that in order for $\widehat{\cQ}$ to satisfy the hermiticity relation~(\ref{eq:BM3}) 
$\Qev$ must inevitably take different forms on GSO($\pm$) sectors. In fact, we find from~(\ref{eq:BE}) 
\begin{eqnarray}
\cR_{\pi}\circ\cQ_{\mathrm{even}}^{\mathrm{GSO}(+)}&=&\frac{q_1}{2i\varepsilon_r}\cR_{\pi}\circ
(\gamma(i)-\gamma(-i)) \nonumber \\
&=&\frac{q_1}{2i\varepsilon_r}\left( (\cR_{\pi}^{\prime}(i))^{-\frac{1}{2}}\gamma(\cR_{\pi}(i))-
(\cR_{\pi}^{\prime}(-i))^{-\frac{1}{2}}\gamma(\cR_{\pi}(-i))\right) \nonumber \\
&=&\frac{q_1}{2\varepsilon_r}(\gamma(i)+\gamma(-i)) \label{eq:BQ}
\end{eqnarray}
because $(\cR_{\pi}^{\prime}(\pm i))^{-\frac{1}{2}}=\pm i$ due to the definition~(\ref{eq:BJ}). 
Thus, we see that the hermiticity condition~(\ref{eq:BP}) is satisfied by our choice~(\ref{eq:BE}) and (\ref{eq:BG}) 
of kinetic operator if we set $q_1=q_2$. With this choice, one can verify that eq.(\ref{eq:BM2}) also holds true. 

As shown above, the ratio $q_1/q_2$ of the finite normalization constants of $\Qev^{\mathrm{GSO}(\pm)}$ has been fixed 
by requiring the hermiticity condition. Then, how can we determine the value of $q_1$ itself? 
This question seems difficult to answer because it requires the detailed information about the reparametrization, 
by which the kinetic operator around the tachyon vacuum has been brought to the simple purely ghost form, 
and the functions  $a_{c,\gamma}(\sigma)$ appearing in~(\ref{eq:IC}). 
We only note here that the sign of $q_1$, which corresponds to the relative sign between the normalization 
constant of $\Qod$ and that of $\Qev$, is related to the choice of the tachyon vacuum around which 
vacuum superstring field theory is constructed. This fact can immediately be seen from the definition~(\ref{eq:AO}) 
of $\Qe$ whose sign is flipped under $A_{0-}\to -A_{0-}$. Since two degenerate tachyon vacua are considered to 
be physically equivalent, we may take $q_1$ to be positive without loss of generality. 

To summarize, we have seen that the twist invariance condition~(\ref{eq:BC}) combined with the hermiticity 
condition~(\ref{eq:BK}) points to the choice 
\begin{eqnarray}
\cQ_{\mathrm{even}}^{\mathrm{GSO}(+)}&=&\frac{q_1}{2i\varepsilon_r}(\gamma(i)-\gamma(-i)) \label{eq:BR} \\
\cQ_{\mathrm{even}}^{\mathrm{GSO}(-)}&=&\frac{q_1}{2\varepsilon_r}(\gamma(i)+\gamma(-i)), \nonumber 
\end{eqnarray}
or collectively 
\begin{equation}
\Qev |\psi\rangle=q_1\frac{1-i}{4\varepsilon_r}(1-i(-1)^{\mathrm{GSO}(\psi)})\left(\gamma(i)-
(-1)^{\mathrm{GSO}(\psi)}\gamma(-i)\right)|\psi\rangle. \label{eq:BS}
\end{equation}
The next step is to check whether the 
above choice of $\Qev$ together with $\Qod$ given in~(\ref{eq:BA}) satisfies the axioms imposed on the 
kinetic operator of superstring field theory. Some of the proofs given below 
overlap with the ones presented in~\cite{ABG}. 
\medskip

\noindent \underline{\textbf{Nilpotency of $\widehat{\cQ}$}} \\
To get a nilpotent kinetic operator $\widehat{\cQ}$, it turns out that 
we must add to $\Qod$ a non-leading term in $\varepsilon_r$ as\footnote{The choice of the second term 
in~(\ref{eq:BT}) is not unique. For example, $\oint\frac{dz}{2\pi i}f(z)b\gamma^2(z)$ will do if a scalar 
function $f(z)$ is regular in an annular region around $|z|=1$ and satisfies $f(\pm i)=1$. Or, we can add to 
$\Qod$ more terms which anticommute with $c(\pm i)$ and $\Qabc$ and are nilpotent themselves. Here 
we consider the simplest choice $\Qabc$.}~\cite{ABG}
\begin{equation}
\Qod =\frac{1}{2i\varepsilon_r^2}(c(i)-c(-i))+\frac{q_1^2}{2}\oint\frac{dz}{2\pi i}b\gamma^2(z). \label{eq:BT}
\end{equation}
Here $\Qabc \equiv\oint\frac{dz}{2\pi i}b\gamma^2(z)$ is part of the BRST charge $Q_B$, and was considered 
in~\cite{ABC}.  Recently, $\Qabc$ was used to propose the pregeometrical formulation of Berkovits' superstring 
field theory~\cite{pregeo}. Since $\Qabc$ is the zero mode of a weight 1 primary $b\gamma^2$, it manifestly 
preserves the twist eigenvalues. Since we have 
\begin{equation}
\widehat{\cQ}\widehat{\cQ}|\widehat{A}\rangle=\left\{(\Qod^2-\Qev^2)\otimes\mathbf{1}-
[\Qod ,\Qev ]\otimes\sigma_1\right\}|\widehat{A}\rangle, 
\end{equation}
we must show both $(\Qod^2-\Qev^2)|\widehat{A}\rangle=0$ and 
\begin{equation}
[\Qod ,\Qev ]|A_{\pm}\rangle=(\Qod\Qev^{\mathrm{GSO}(\pm)}-\Qev^{\mathrm{GSO}(\pm)}\Qod)|A_{\pm}\rangle=0. \label{eq:BTa}
\end{equation}
The latter holds because $\Qod$ (\ref{eq:BT}) 
contains no $\beta$ field \textit{and $\Qod$ does not change the GSO parity}, as 
indicated in~(\ref{eq:BTa}).\footnote{Note that, if we had implemented the grading in~(\ref{eq:BS}) by 
the Grassmannality or the ghost number of the states $|\psi\rangle$, then $\Qev$ would not commute with $\Qod$ 
because $\Qod$ itself is Grassmann-odd and has ghost number 1. So the grading had to be implemented by the GSO 
parity for $\widehat{\cQ}$ to be nilpotent.} The former one can be shown as follows: 
\begin{equation}
\Qod^2|A_{\pm}\rangle=\frac{q_1^2}{4i\varepsilon_r^2}\left\{\oint\frac{dz}{2\pi i}b\gamma^2(z), c(i)-c(-i)\right\}
|A_{\pm}\rangle=\frac{q_1^2}{4i\varepsilon_r^2}\left(\gamma(i)^2-\gamma(-i)^2\right)|A_{\pm}\rangle, \label{eq:BU}
\end{equation}
and, from eqs.(\ref{eq:BR}), 
\begin{equation}
\Qev^2|A_{\pm}\rangle=\Qev^{\mathrm{GSO}(\mp)}\Qev^{\mathrm{GSO}(\pm)}|A_{\pm}\rangle=\frac{q_1^2}{4i\varepsilon_r^2}
\left(\gamma(i)^2-\gamma(-i)^2\right)|A_{\pm}\rangle, \label{eq:BV}
\end{equation}
where $|A_{+/-}\rangle$ denote any states in the GSO($+/-$) sectors respectively, and we have used the fact that 
$\Qev$ reverses the GSO parity of the states. From~(\ref{eq:BU}) and (\ref{eq:BV}), it follows that 
$(\Qod^2-\Qev^2)|\widehat{A}\rangle=0$. This completes the proof that 
$\widehat{\cQ}=\Qod\otimes\sigma_3-\Qev\otimes i\sigma_2$ with $\Qod$ given in~(\ref{eq:BT}) and 
$\Qev$ in~(\ref{eq:BR}) is nilpotent. 
\medskip

\noindent \underline{$\langle\cI|\widehat{\cQ}=0$} \\
Given that the identity string field $\langle\cI |$ is defined as 
\begin{equation}
\langle\cI |\varphi\rangle =\langle f^{(1)}_1\circ\varphi (0)\rangle_{\mathrm{UHP}} \label{eq:BW}
\end{equation}
with $f^{(1)}_1(z)=h^{-1}(h(z)^2)=\frac{2z}{1-z^2}$ for an arbitrary Fock space state 
$|\varphi\rangle =\varphi(0)|0\rangle$, both $\langle\cI |c(\pm i)$ and $\langle\cI |\gamma (\pm i)$ 
contain divergences because the conformal factors $(f^{(1)\prime}_1(\pm i))^h$ diverge for $h<0$. 
However, as pointed out in~\cite{GRSZ1,ABG}, they can be regularized by the following prescription. 
If we make the replacements 
\begin{eqnarray}
c(i)&\longrightarrow& c_{\epsilon}(i)=\frac{1}{2}\left( e^{-i\epsilon}c(ie^{i\epsilon})+e^{i\epsilon}
c(ie^{-i\epsilon})\right), \nonumber \\
c(-i)&\longrightarrow& c_{\epsilon}(-i)=\frac{1}{2}\left( e^{-i\epsilon}c(-ie^{i\epsilon})+e^{i\epsilon}
c(-ie^{-i\epsilon})\right), \label{eq:BX} \\
\gamma(i)&\longrightarrow& \gamma_{\epsilon}(i)=\frac{1}{e^{-\frac{\pi i}{4}}-e^{\frac{\pi i}{4}}}
\left( e^{-\frac{\pi i}{4}-\frac{i\epsilon}{2}}\gamma(ie^{i\epsilon})-e^{\frac{\pi i}{4}+\frac{i\epsilon}{2}}
\gamma(ie^{-i\epsilon})\right), \nonumber \\
\gamma(-i)&\longrightarrow& \gamma_{\epsilon}(-i)=\frac{1}{e^{-\frac{\pi i}{4}}-e^{\frac{\pi i}{4}}}
\left( e^{-\frac{\pi i}{4}-\frac{i\epsilon}{2}}\gamma(-ie^{i\epsilon})-e^{\frac{\pi i}{4}+\frac{i\epsilon}{2}}
\gamma(-ie^{-i\epsilon})\right), \nonumber
\end{eqnarray}
in $\widehat{\cQ}$, then all of $c_{\epsilon}(\pm i),\gamma_{\epsilon}(\pm i)$ turn out to annihilate $\langle\cI |$, 
while in the $\epsilon\to 0$ limit they na\"{\i}vely reduce to the original midpoint insertions. 
Here we give a slightly different proof than in~\cite{ABG} that $\gamma_{\epsilon}(i)$ kills the identity. Let us consider 
\begin{equation}
\gamma_{(a,b)}(i)=\frac{1}{a+b}\left( a\gamma(ie^{i\epsilon})+b\gamma(ie^{-i\epsilon})\right) \label{eq:BY}
\end{equation}
with $\epsilon$-dependent constants $a$ and $b$, and see when $\langle\cI |\gamma_{(a,b)}(i)|\varphi\rangle$ 
vanishes. Note that in the $\epsilon\to 0$ limit $\gamma_{(a,b)}(i)$ reduces to $\gamma(i)$ irrespective of 
the values of $a$ and $b$. From the definition~(\ref{eq:BW}) of the identity, we have 
\begin{eqnarray}
\langle\cI |\gamma_{(a,b)}(i)|\varphi\rangle&=&\frac{a}{a+b}\left\langle\left( f^{(1)\prime}_1(ie^{i\epsilon})
\right)^{-\frac{1}{2}}\gamma(f^{(1)}_1(ie^{i\epsilon}))f^{(1)}_1\circ\varphi(0)\right\rangle_{\mathrm{UHP}}\nonumber \\
& &\hspace{-1.5cm}{}+\frac{b}{a+b}\left\langle\left( f^{(1)\prime}_1(ie^{-i\epsilon})
\right)^{-\frac{1}{2}}\gamma(f^{(1)}_1(ie^{-i\epsilon}))f^{(1)}_1\circ\varphi(0)\right\rangle_{\mathrm{UHP}}. \label{eq:BZ}
\end{eqnarray}
Using the relations 
\begin{eqnarray*}
f^{(1)}_1(ie^{-i\epsilon})&=&\frac{2ie^{-i\epsilon}}{1+e^{-2i\epsilon}}=\frac{2ie^{i\epsilon}}{e^{2i\epsilon}+1}
=f^{(1)}_1(ie^{i\epsilon}), \\
f^{(1)\prime}_1(ie^{-i\epsilon})&=&\frac{2(1+z^2)}{(1-z^2)^2}\Bigg|_{z=ie^{-i\epsilon}}=\frac{2(1-e^{-2i\epsilon})
}{(1+e^{-2i\epsilon})^2}=e^{2i\epsilon}\frac{2(e^{2i\epsilon}-1)}{(e^{2i\epsilon}+1)^2} \\
&=& e^{\pi i+2i\epsilon}\frac{2(1-e^{2i\epsilon})}{(1+e^{2i\epsilon})^2}=e^{\pi i+2i\epsilon}f^{(1)\prime}_1
(ie^{i\epsilon}), 
\end{eqnarray*}
where we have conventionally defined $-1=e^{\pi i}$, we rewrite eq.(\ref{eq:BZ}) as 
\[ \langle\cI |\gamma_{(a,b)}(i)|\varphi\rangle=\frac{1}{a+b}\left( f^{(1)\prime}_1(ie^{i\epsilon})\right)^{-\frac{1}{2}}
\left( a+be^{-\frac{\pi i}{2}-i\epsilon}\right)\left\langle\gamma(f^{(1)}_1(ie^{i\epsilon}))f^{(1)}_1\circ\varphi(0)
\right\rangle_{\mathrm{UHP}}. \]
This expression vanishes when 
\[ \frac{b}{a}=-e^{\frac{\pi i}{2}+i\epsilon} \]
is satisfied. So, by choosing 
\[ a=e^{-\frac{\pi i}{4}-\frac{i\epsilon}{2}}, \qquad b=-e^{\frac{\pi i}{4}+\frac{i\epsilon}{2}}, \]
$\gamma_{(a,b)}(i)$ (\ref{eq:BY}) essentially reproduces $\gamma_{\epsilon}(i)$ of (\ref{eq:BX}). 
In the same way we can prove $\langle\cI |\gamma_{\epsilon}(-i)=0,\ \langle\cI |c_{\epsilon}(\pm i)=0$ 
as well. Furthermore, $\Qabc$ also annihilates $\langle\cI|$ because 
\[ \langle\cI|\Qabc|\varphi\rangle=\left\langle f^{(1)}_1\circ\left(\oint\frac{dz}{2\pi i}b\gamma^2(z)
\varphi(0)\right)\right\rangle_{\mathrm{UHP}}=\left\langle \oint\frac{dz^{\prime}}{2\pi i}b\gamma^2(z^{\prime})
\left(f^{(1)}_1\circ\varphi(0)\right)\right\rangle_{\mathrm{UHP}} \]
is shown to vanish by the contour deformation argument. 
Therefore, if we define $\Qod$ and $\Qev^{\mathrm{GSO}(\pm)}$ as the $\epsilon\to 0$ limit of 
\begin{eqnarray*}
\cQ_{\mathrm{odd},\epsilon}&=&\frac{1}{2i\varepsilon_r^2}\left( c_{\epsilon}(i)-c_{\epsilon}(-i)\right)
+\frac{q_1^2}{2}\oint\frac{dz}{2\pi i}b\gamma^2(z), \\
\cQ_{\mathrm{even},\epsilon}^{\mathrm{GSO}(+)}&=&\frac{q_1}{2i\varepsilon_r}
\left( \gamma_{\epsilon}(i)-\gamma_{\epsilon}(-i)\right), \\
\cQ_{\mathrm{even},\epsilon}^{\mathrm{GSO}(-)}&=&\frac{q_1}{2\varepsilon_r}
\left( \gamma_{\epsilon}(i)+\gamma_{\epsilon}(-i)\right), 
\end{eqnarray*}
respectively, then we obtain an operator $\widehat{\cQ}$ which annihilates the identity. 
\smallskip

To give still another argument that $\Qev$ kills $|\cI\rangle$, we notice that the action of $\Qev$ on 
a state $|\psi\rangle$ can be expressed\footnote{As was shown in~\cite{GRSZ1}, almost the same is true of $\Qod$: 
see eq.(\ref{eq:DE}).} as 
an inner derivation,\footnote{Note that this expression has the same structure as 
the original definition~(\ref{eq:AO}) of $\Qe$. Since the $\epsilon\to 0$ limit of $\SIG$ is not well-defined, 
$\Qev$ itself may not be considered as an inner derivation.} 
\begin{eqnarray}
\Qev |\psi\rangle&=&\lim_{\epsilon\to 0}\left( |\SIG *\psi\rangle -(-1)^{\mathrm{GSO}(\psi)}|\psi *
\SIG \rangle\right), \nonumber \\
|\SIG\rangle&=&\Gamma_\epsilon|\cI\rangle, \label{eq:CH} \\
\Gamma_{\epsilon}&=&q_1\frac{1-i}{4\varepsilon_r}\left(\gamma(ie^{i\epsilon})+i\gamma(-ie^{-i\epsilon})\right). \nonumber 
\end{eqnarray}
As shown in Appendix~\ref{sec:appA}, by considering the inner product $\langle\varphi |\Qev |\psi\rangle$ with a Fock space state 
$\langle\varphi |$ we actually recover the previous expression~(\ref{eq:BS}) 
\begin{equation}
\langle\varphi |\Qev |\psi\rangle=q_1\frac{1-i}{4\varepsilon_r}\left(1-i(-1)^{\mathrm{GSO}(\psi)}\right)
\langle\varphi |\left(\gamma(i)-(-1)^{\mathrm{GSO}(\psi)}\gamma(-i)\right)|\psi\rangle. \label{eq:CI}
\end{equation}
From the expression~(\ref{eq:CH}), it is obvious that $\Qev$ annihilates the identity $|\cI\rangle$. 
Substituting $|\psi\rangle=|\cI\rangle$ and $(-1)^{\mathrm{GSO}(\cI)}=+1$, one finds 
\[ \Qev |\cI\rangle=\lim_{\epsilon\to 0}(|\SIG *\cI\rangle -|\cI *\SIG\rangle)=\lim_{\epsilon\to 0}
(|\SIG\rangle -|\SIG\rangle)=0. \]
\medskip

\noindent \underline{\textbf{Derivation property of} $\widehat{\cQ}$} \\
It is known~\cite{GRSZ1} that $\Qmid =\frac{1}{2i}(c(i)-c(-i))$ is a graded derivation of the $*$-algebra 
because $\Qmid$ can be written as 
\begin{eqnarray*}
\Qmid&=&\sum_{n=0}^{\infty}(-1)^n\cC_{2n}, \\
& &\hspace{-5mm} \cC_0=c_0, \qquad \cC_n=c_n+(-1)^nc_{-n} \quad \mbox{for } n\neq 0, 
\end{eqnarray*}
and each $\cC_n$ obeys the Leibniz rule graded by the Grassmannality~\cite{RZ,RSZ1}. 
The derivation property of $\Qabc$, which is the zero-mode of a primary field of conformal weight 1, 
is proven by  the contour deformation argument~\cite{RZ}. 
Taking the internal Chan-Paton 
factors into account, $\widehat{\cQ}_{\mathrm{odd}}=\Qod\otimes\sigma_3$ satisfies 
\begin{equation}
\widehat{\cQ}_{\mathrm{odd}}(\widehat{A}*\widehat{B})=(\widehat{\cQ}_{\mathrm{odd}}\widehat{A})*\widehat{B}
+(-1)^{\gh (\widehat{A})}\widehat{A}*(\widehat{\cQ}_{\mathrm{odd}}\widehat{B}), \label{eq:CF}
\end{equation}
where the internal Chan-Paton structure of $\widehat{A}$ and $\widehat{B}$ is given by (\ref{eq:BL}). 
For the case of $\Qev$, we will make use of the expression~(\ref{eq:CH}). Let us consider $\Qev$ acting 
on the $*$-product $A*B$ of two states $A$ and $B$. From the property of the GSO parity that 
$(-1)^{\mathrm{GSO}(A*B)}=(-1)^{\mathrm{GSO}(A)}(-1)^{\mathrm{GSO}(B)}$ one obtains 
\begin{eqnarray*}
\Qev |A*B\rangle &=& |\SIG *A*B\rangle-(-1)^{\mathrm{GSO}(A*B)}|A*B*\SIG\rangle \\
&=& \left( |\SIG *A\rangle-(-1)^{\mathrm{GSO}(A)}|A*\SIG\rangle\right)*|B\rangle \\ & &
{}+(-1)^{\mathrm{GSO}(A)}|A\rangle *\left( |\SIG *B\rangle-(-1)^{\mathrm{GSO}(B)}|B*\SIG\rangle\right) \\
&=& |(\Qev A)*B\rangle +(-1)^{\mathrm{GSO}(A)}|A*(\Qev B)\rangle,
\end{eqnarray*}
where we have omitted the symbol $\lim_{\epsilon\to 0}$. Attaching the Chan-Paton factors to $A$ and $B$, and 
then multiplying $i\sigma_2$ from the left, we have for $\widehat{\cQ}_{\mathrm{even}}=\Qev\otimes i\sigma_2$ 
\begin{equation}
\widehat{\cQ}_{\mathrm{even}}|\widehat{A}*\widehat{B}\rangle=|(\widehat{\cQ}_{\mathrm{even}}\widehat{A})*
\widehat{B}\rangle +(-1)^{\mathrm{GSO}(\widehat{A})}
i\sigma_2 |\widehat{A}*(\Qev\widehat{B})\rangle. \label{eq:CJ}
\end{equation}
When $i\sigma_2$ passes $\widehat{A}$, we find from (\ref{eq:BL}) a rule 
\begin{eqnarray*}
i\sigma_2\cdot\widehat{A}=-(-1)^{\mathrm{GSO}(\widehat{A})}\widehat{A}\cdot i\sigma_2 & & \mbox{ for } \gh (\widehat{A}) 
\mbox{ odd,} \\
i\sigma_2\cdot\widehat{A}=(-1)^{\mathrm{GSO}(\widehat{A})}\widehat{A}\cdot i\sigma_2 & & \mbox{ for } \gh (\widehat{A}) 
\mbox{ even,}
\end{eqnarray*}
which can be written collectively in the form 
\[ i\sigma_2\cdot\widehat{A}=(-1)^{\gh (\widehat{A})}(-1)^{\mathrm{GSO}(\widehat{A})}\widehat{A}\cdot i\sigma_2. \]
Applying it to eq.(\ref{eq:CJ}), we eventually find 
\begin{equation}
\widehat{\cQ}_{\mathrm{even}}(\widehat{A}*\widehat{B})=(\widehat{\cQ}_{\mathrm{even}}\widehat{A})*\widehat{B}
+(-1)^{\gh (\widehat{A})}\widehat{A}*(\widehat{\cQ}_{\mathrm{even}}\widehat{B}). \label{eq:CK}
\end{equation}
Since both $\widehat{\cQ}_{\mathrm{odd}}$ and $\widehat{\cQ}_{\mathrm{even}}$ obey the same Leibniz rule~(\ref{eq:CF}), 
(\ref{eq:CK}) with the ghost number grading, so does 
$\widehat{\cQ}=\widehat{\cQ}_{\mathrm{odd}}-\widehat{\cQ}_{\mathrm{even}}$: 
\begin{equation}
\widehat{\cQ}(\widehat{A}*\widehat{B})=(\widehat{\cQ}\widehat{A})*\widehat{B}
+(-1)^{\gh (\widehat{A})}\widehat{A}*(\widehat{\cQ}\widehat{B}). \label{eq:CL}
\end{equation}
\medskip

\noindent \underline{\textbf{Hermiticity of $\widehat{\cQ}$ in the presence of $Y_{-2}$}} \\
As mentioned before, 
in the proof of gauge invariance of the action~(\ref{eq:CA}) we are going to use the hermiticity relations 
\begin{equation}
\llk\widehat{Y}_{-2}|\widehat{\cQ}\widehat{A},\widehat{B}\rrk =-(-1)^{\gh(\widehat{A})}\llk\widehat{Y}_{-2}|
\widehat{A},\widehat{\cQ}\widehat{B}\rrk, \label{eq:DA}
\end{equation}
or more generally 
\begin{equation}
\llk\widehat{Y}_{-2}|\widehat{\cQ}\widehat{A}_1,\widehat{A}_2*\ldots *\widehat{A}_n\rrk =-(-1)^{\gh(\widehat{A}_1)}
\llk\widehat{Y}_{-2}|\widehat{A}_1,\widehat{\cQ}(\widehat{A}_2*\ldots *\widehat{A}_n)\rrk. \label{eq:DB}
\end{equation}
To show them we decompose $\widehat{\cQ},\widehat{A},\widehat{B}$ into their GSO components. 
For the first one~(\ref{eq:DA}) explicit forms are 
\begin{eqnarray}
\llk Y_{-2}|\Qod A_+,B_+\rrk&=&-(-1)^{|A_+|}\llk Y_{-2}|A_+,\Qod B_+\rrk, \label{eq:DC1}\\
\llk Y_{-2}|\Qod A_-,B_-\rrk&=&-(-1)^{|A_-|}\llk Y_{-2}|A_-,\Qod B_-\rrk, \label{eq:DC2}\\
\llk Y_{-2}|\Qev A_+,B_-\rrk&=&-\llk Y_{-2}|A_+,\Qev B_-\rrk, \label{eq:DC3} \\
\llk Y_{-2}|\Qev A_-,B_+\rrk&=& \llk Y_{-2}|A_-,\Qev B_+\rrk. \label{eq:DC4} 
\end{eqnarray}
(\ref{eq:DC3}), (\ref{eq:DC4}) have already been proved below eq.(\ref{eq:BM3}). We can show that 
relations~(\ref{eq:DC1}), (\ref{eq:DC2}) are satisfied by $\Qmid$ from the precise definition of the 2-vertex, 
\begin{equation}
\llk Y_{-2}|A,B\rrk=\langle Y(i)Y(-i)A(0)\cR_{\pi}\circ B(0)\rangle_{\mathrm{UHP}}, \label{eq:DD}
\end{equation}
and the conformal transformation of $\Qmid$ under $\cR_{\pi}$, 
\[ \cR_{\pi}\circ\Qmid=-\Qmid. \]
For $\Qabc$, eqs.(\ref{eq:DC1}), (\ref{eq:DC2}) are equivalent to the statement 
\begin{equation}
\left\langle Y(i)Y(-i)\oint_{C}\frac{dz}{2\pi i}b\gamma^2(z)(A_{\pm}(0)\ \cR_{\pi}\circ B_{\pm}(0))
\right\rangle_{\mathrm{UHP}}=0, \label{eq:IA}
\end{equation}
where $C$ is an integration contour encircling $0$ and $\cR_{\pi}(0)=\infty$, but not $\pm i$. 
By deforming the contour this equation is restated as 
\begin{equation}
[\Qabc,Y_{-2}]\equiv\left(\oint_{C(i)}\frac{dz}{2\pi i}b\gamma^2(z)Y(i)\right)Y(-i)+
\left(\oint_{C(-i)}\frac{dz}{2\pi i}b\gamma^2(z)Y(-i)\right)Y(i)=0, \label{eq:IB}
\end{equation}
where $C(\pm i)$ are small contours encircling $\pm i$, respectively. This equality holds because the 
operator product between $b\gamma^2=b\eta\partial\eta e^{2\phi}$ and $Y=c\partial\xi e^{-2\phi}$ is 
non-singular. Thus we have shown that $\widehat{\cQ}$ satisfies the hermiticity relation~(\ref{eq:DA}). 

In the more general case~(\ref{eq:DB}), we make use of the inner derivation formulas: eq.(\ref{eq:CH}) 
for $\Qev$ and 
\begin{eqnarray}
\Qmid |\psi\rangle&=& \lim_{\epsilon\to 0}\left(|S_{\epsilon}*\psi\rangle -(-1)^{|\psi|}|\psi *S_{\epsilon}\rangle
\right), \label{eq:DE} \\
|S_{\epsilon}\rangle&=&\frac{1}{4i}\left( e^{-i\epsilon}c(ie^{i\epsilon})-e^{i\epsilon}c(-i
e^{-i\epsilon})\right) |\cI\rangle, \nonumber 
\end{eqnarray}
for $\Qmid$~\cite{GRSZ1,KisOhm}. We find 
\begin{eqnarray}
& &\llk Y_{-2}|\Qmid A_{1+},(\widehat{A}_2*\ldots *\widehat{A}_n)_+\rrk =\lim_{\epsilon\to 0}\biggl(\llk Y_{-2}|
S_{\epsilon}*A_{1+},(\widehat{A}_2* \ldots*\widehat{A}_n)_+\rrk \nonumber \\
& &\hspace{6.5cm} {}-(-1)^{|A_{1+}|}\llk Y_{-2}|A_{1+}*S_{\epsilon},(\widehat{A}_2* \ldots*\widehat{A}_n)_+\rrk
\biggr) \nonumber \\ 
& &\hspace{1cm}= -(-1)^{|A_{1+}|}\lim_{\epsilon\to 0}\biggl(\llk Y_{-2}|A_{1+},S_{\epsilon}*(\widehat{A}_2* \ldots*\widehat{A}_n)_+
\rrk \nonumber \\
& &\qquad\qquad\quad {}-(-1)^{|(\widehat{A}_2* \ldots*\widehat{A}_n)_+|}\llk Y_{-2}|A_{1+},(\widehat{A}_2* \ldots*\widehat{A}_n)_+
*S_{\epsilon}\rrk\biggr) \nonumber \\
& &\hspace{1cm}= -(-1)^{|A_{1+}|}\llk Y_{-2}|A_{1+},\Qmid (\widehat{A}_2* \ldots*\widehat{A}_n)_+\rrk , \label{eq:DF}
\end{eqnarray}
and similarly 
\begin{equation}
\llk Y_{-2}|\Qmid A_{1-},(\widehat{A}_2* \ldots*\widehat{A}_n)_-\rrk=-(-1)^{|A_{1-}|}
\llk Y_{-2}|A_{1-}, \Qmid (\widehat{A}_2* \ldots*\widehat{A}_n)_-\rrk, \label{eq:DG}
\end{equation}
where we have used the associativity of the $*$-product and cyclicity for a GSO($+$) state 
$S_{\epsilon}$, and $(\widehat{A}_2* \ldots*\widehat{A}_n)_{+/-}$ denote the GSO($+/-$) components of 
$\widehat{A}_2* \ldots*\widehat{A}_n$. Note that $(-1)^{|A_1|}$ and $(-1)^{|A_2*\ldots *A_n|}$ agree with 
each other since the whole insertion inside the correlator must be 
Grassmann-odd to give a non-vanishing value. In the same way as above, we get 
\begin{eqnarray}
& &\hspace{-1.2cm}\llk Y_{-2}|\Qev A_{1+},(\widehat{A}_2* \ldots*\widehat{A}_n)_-\rrk =\lim_{\epsilon\to 0}
\biggl(\llk Y_{-2}|\SIG *A_{1+},(\widehat{A}_2* \ldots*\widehat{A}_n)_-\rrk\nonumber \\
& &\hspace{6.5cm} {}-(-1)^{\mathrm{GSO}(A_{1+})}\llk Y_{-2}|A_{1+}*\SIG ,(\widehat{A}_2* \ldots*\widehat{A}_n)_-\rrk
\biggr)\nonumber \\
&=&-\lim_{\epsilon\to 0}\left(\llk Y_{-2}|A_{1+},\SIG *(\widehat{A}_2* \ldots*\widehat{A}_n)_-\rrk-(-1)
\llk Y_{-2}|A_{1+}, (\widehat{A}_2* \ldots*\widehat{A}_n)_-*\SIG\rrk\right) \nonumber \\
&=& -\llk Y_{-2}|A_{1+},\Qev (\widehat{A}_2* \ldots*\widehat{A}_n)_-\rrk, \label{eq:DH}
\end{eqnarray}
and 
\begin{equation}
\llk Y_{-2}|\Qev A_{1-},(\widehat{A}_2* \ldots*\widehat{A}_n)_+\rrk =\llk Y_{-2}|A_{1-},
\Qev (\widehat{A}_2* \ldots*\widehat{A}_n)_-\rrk. \label{eq:DI}
\end{equation}
In the second equality of (\ref{eq:DH}), an additional sign factor $(-1)$ has arisen because $\SIG$ lives in 
the GSO($-$) sector (see eq.(\ref{eq:AI})). The proof of~(\ref{eq:DB}) for $\Qabc$ is essentially the same 
as in the previous case~(\ref{eq:IA}).
Collecting eqs.(\ref{eq:DF})--(\ref{eq:DI}) in the 
matrix form gives eq.(\ref{eq:DB}).

\subsection{Gauge invariant cubic action of vacuum superstring field theory}\label{subsec:gauge}
In this subsection we will show that the following cubic action 
\begin{equation}
S_V=-\frac{\kappa_0}{2}\mathrm{Tr}\left[\frac{1}{2}\llk\widehat{Y}_{-2}|\widehat{\cA},\widehat{\cQ}
\widehat{\cA}\rrk +\frac{1}{3}\llk\widehat{Y}_{-2}|\widehat{\cA},\widehat{\cA}*\widehat{\cA}\rrk\right] \label{eq:CA}
\end{equation}
with $\widehat{\cQ}$ given above (eqs.(\ref{eq:BT}), (\ref{eq:BR})) is invariant under the gauge transformation 
\begin{equation}
\delta\widehat{\cA}=\widehat{\cQ}\widehat{\Lambda}+\widehat{\cA}*\widehat{\Lambda}-\widehat{\Lambda}*
\widehat{\cA}, \label{eq:CB}
\end{equation}
where $\widehat{\Lambda}$ is an infinitesimal gauge parameter of ghost number 0 and picture number 0, whose 
internal Chan-Paton structure is 
\begin{equation}
\widehat{\Lambda}=\Lambda_+\otimes\mathbf{1}+\Lambda_-\otimes\sigma_1. \label{eq:CC}
\end{equation}
The gauge variation of the action linear in $\widehat{\Lambda}$ is 
\begin{eqnarray}
\delta S_V&=&-\frac{\kappa_0}{2}\mathrm{Tr}\left[\llk\widehat{Y}_{-2}|\widehat{\cA},\widehat{\cQ}(\delta\widehat{\cA})
\rrk+\llk\widehat{Y}_{-2}|\delta\widehat{\cA},\widehat{\cA}*\widehat{\cA}\rrk\right] \label{eq:CD} \\
&=&-\frac{\kappa_0}{2}\mathrm{Tr}\Biggl[\llk\widehat{Y}_{-2}|\widehat{\cA},\widehat{\cQ}^2\widehat{\Lambda}\rrk
+\llk\widehat{Y}_{-2}|\widehat{\cA},\widehat{\cQ}(\widehat{\cA}*\widehat{\Lambda})\rrk -\llk\widehat{Y}_{-2}|
\widehat{\cA},\widehat{\cQ}(\widehat{\Lambda}*\widehat{\cA})\rrk \nonumber \\
& &{}+\llk\widehat{Y}_{-2}|\widehat{\cQ}\widehat{\Lambda},\widehat{\cA}*\widehat{\cA}\rrk +\llk\widehat{Y}_{-2}|
(\widehat{\cA}*\widehat{\Lambda}),\widehat{\cA}*\widehat{\cA}\rrk -\llk\widehat{Y}_{-2}|(\widehat{\Lambda}
*\widehat{\cA}), \widehat{\cA}*\widehat{\cA}\rrk \Biggr], \nonumber 
\end{eqnarray}
where we have used the cyclicity of the vertices and the hermiticity (\ref{eq:DA}) of $\widehat{\cQ}$. 
Making use of the cyclicity and the associativity of the $*$-product, the last two terms cancel each other. 
The first term vanishes as well thanks to the nilpotency of $\widehat{\cQ}$. Using the cyclicity, 
the `partial integration'~(\ref{eq:DB}) and the derivation property~(\ref{eq:CL}) 
of $\widehat{\cQ}$, the second and the third terms are written as 
\begin{eqnarray*}
& &\hspace{-1cm}
\mathrm{Tr}\left(\llk\widehat{Y}_{-2}|\widehat{\cQ}\widehat{\cA},\widehat{\cA}*\widehat{\Lambda}\rrk -
\llk\widehat{Y}_{-2}|\widehat{\cQ}\widehat{\cA},\widehat{\Lambda}*\widehat{\cA}\rrk \right) \\
&=&\mathrm{Tr}\left(\llk\widehat{Y}_{-2}|\widehat{\Lambda},\widehat{\cQ}\widehat{\cA}*\widehat{\cA}\rrk
-\llk\widehat{Y}_{-2}|\widehat{\Lambda},\widehat{\cA}*\widehat{\cQ}\widehat{\cA}\rrk\right) \\
&=&\mathrm{Tr}\llk\widehat{Y}_{-2}|\widehat{\Lambda},\widehat{\cQ}(\widehat{\cA}*\widehat{\cA})\rrk
=-\mathrm{Tr}\llk\widehat{Y}_{-2}|\widehat{\cQ}\widehat{\Lambda},\widehat{\cA}*\widehat{\cA}\rrk, 
\end{eqnarray*}
which precisely cancels the fourth term. This completes the proof of gauge invariance. 
Therefore the action~(\ref{eq:CA}) defines a consistent gauge theory of a string field, at least at the 
classical level. Moreover, the kinetic operator $\widehat{\cQ}$, which governs the perturbative nature 
of the string field around $\widehat{\cA}=0$, has the following properties: 
\begin{enumerate}
  \item $\widehat{\cQ}$ is made purely out of ghost operators; 
  \item $\widehat{\cQ}$ has vanishing cohomology, so that there are no perturbative physical open string 
  states\footnote{We are ignoring the effect of the non-trivial kernel of $Y_{-2}$ on the physical 
  open string spectrum~\cite{Berkovits}.} around the vacuum $\widehat{\cA}=0$; 
  \item $\widehat{\cQ}$ contains non-zero $\Qev$ so that the unwanted $\zetto_2$ reflection symmetry 
  is broken, as it should be;
  \item $\widehat{\cQ}$ has been constructed in such a way that the action $S_V$~(\ref{eq:CA}) is 
  invariant under the twist transformation~(\ref{eq:AZ}). 
\end{enumerate}
The property 1 is obvious from the explicit form~(\ref{eq:BT}), (\ref{eq:BR}) of $\widehat{\cQ}$. 
To show the property 2, suppose that we have a state $|\widehat{\cA}\rangle$ which is annihilated by 
$\widehat{\cQ}$. Then $|\widehat{\cA}\rangle$ itself can be written as 
\[ |\widehat{\cA}\rangle=\{\widehat{\cQ},\varepsilon_r^2\widehat{b}_0\}|\widehat{\cA}\rangle=\widehat{\cQ}
(\varepsilon_r^2\widehat{b}_0|\widehat{\cA}\rangle ), \]
where $\widehat{b}_0=b_0\otimes\sigma_3$. Since any $\widehat{\cQ}$-closed state $|\widehat{\cA}\rangle$ 
has, at least formally, been expressed as a $\widehat{\cQ}$-exact form, it follows that $\widehat{\cQ}$ has vanishing 
cohomology. 
Twist invariance of the action is explicitly shown in Appendix~\ref{sec:appB}. 
From the properties mentioned above, it seems that 
the action~(\ref{eq:CA}) is well suited for the description of cubic superstring field theory around 
one of the tachyon vacua. Although this action is gauge-invariant and is well-defined even for 
finite $\varepsilon_r$, if we are to consider that 
the vacuum superstring field theory action~(\ref{eq:CA}) is derived from the original one~(\ref{eq:AA}) 
through the tachyon condensation followed by a field redefinition, the parameter $\varepsilon_r$ should be 
taken to zero. 

\sectiono{Construction of a Spacetime-Independent Solution: $\varepsilon_r$-Expansion}\label{sec:solution}
Now that we have determined the precise form of the cubic vacuum superstring field theory action, 
we set about constructing solutions in this theory. Varying the action~(\ref{eq:CA}), we obtain 
the following equations of motion: 
\begin{equation}
\cF(\widehat{\cA})\equiv\widehat{\cQ}\widehat{\cA}+\widehat{\cA}*\widehat{\cA}=0 \label{eq:EB}
\end{equation}
or, in the GSO-component form $\cF(\widehat{\cA})=\cF_+\otimes\mathbf{1}+\cF_-\otimes\sigma_1$, 
\begin{eqnarray}
\cF_+&\equiv&\Qod\cA_++\cA_+*\cA_+ +\Qev\cA_- -\cA_- *\cA_-=0, \label{eq:EC} \\
\cF_-&\equiv&\Qod\cA_- +\Qev\cA_+ +\cA_+ *\cA_- -\cA_- *\cA_+ =0. \label{eq:ED}
\end{eqnarray}
It has been discussed in the literature~\cite{RSZ6,HM,RV1,Okawa} that, even if the inner product 
$\langle\widehat{\varphi}|\cF(\widehat{\cA})\rangle$ with any Fock space state $\langle\widehat{\varphi}|$ 
vanishes, it does \textit{not} follow that $\langle\widehat{X}|\cF(\widehat{\cA})\rangle$ vanishes for 
more general states $\langle\widehat{X}|$. However, since it seems that what matters in this argument 
is the normalization of the solution $\widehat{\cA}$ and we are not in a position to deal with it 
in detail, we consider the simplest case where we require 
\begin{equation}
\langle\widehat{\varphi}|\cF(\widehat{\cA})\rangle =\langle\widehat{\varphi}|\widehat{\cQ}\widehat{\cA}
+\widehat{\cA}*\widehat{\cA}\rangle =0 \label{eq:EE}
\end{equation}
to hold for any Fock space state $\langle\widehat{\varphi}|$ of ghost number 1 and \textit{picture number}
$\mathit{-2}$, instead of inserting the double-step inverse picture changing operator $Y_{-2}$. 
In the component form, it becomes 
\begin{equation}
\langle\varphi_+|\cF_+\rangle =\langle\varphi_-|\cF_-\rangle =0, \label{eq:EF}
\end{equation}
where $\varphi_+,\varphi_-$ denote the GSO($+$)-, GSO($-$)-components of 
$\widehat{\varphi}=\varphi_+\otimes\sigma_3+\varphi_-\otimes i\sigma_2$, respectively. 
\medskip

The fact that we must handle two quite different types of operators 
$\Qmid=\frac{1}{2i}(c(i)-c(-i))$, $\Qabc=\oint\frac{dz}{2\pi i}b\gamma^2(z)$ at the same time makes it 
difficult to find solutions of the equations of motion. Now let us recall that in the `realistic' vacuum 
superstring field theory the parameter $\varepsilon_r$ has to be taken to zero and that 
$\Qmid$ and $\Qabc$ enter $\Qod$ in different orders in $\varepsilon_r$. 
Multiplied by $\varepsilon_r^4$, the equations of motion~(\ref{eq:EF}) can be written as 
\begin{eqnarray}
& &\langle\varphi_+|\left(\Qmid +\frac{q_1^2\varepsilon_r^2}{2}\Qabc\right)(\varepsilon_r^2\cA_+)+
(\varepsilon_r^2\cA_+)*(\varepsilon_r^2\cA_+) \nonumber \\
& &\hspace{2cm} +\varepsilon_r\Gamma(\varepsilon_r^2\cA_-)-(\varepsilon_r^2\cA_-)*
(\varepsilon_r^2\cA_-)\rangle =0, \label{eq:EG} \\
& &\langle\varphi_-|\left(\Qmid +\frac{q_1^2\varepsilon_r^2}{2}\Qabc\right)(\varepsilon_r^2\cA_-)
+\varepsilon_r\Gamma(\varepsilon_r^2\cA_+) \nonumber \\
& &\hspace{2cm} +(\varepsilon_r^2\cA_+)*(\varepsilon_r^2\cA_-)-(\varepsilon_r^2\cA_-)*
(\varepsilon_r^2\cA_+)\rangle =0, \label{eq:EH}
\end{eqnarray}
where we have introduced the notation $\Gamma=\varepsilon_r \Qev$ which is finite in the 
$\varepsilon_r\to 0$ limit. If we assume that $\cA^{\prime}_{\pm}\equiv\varepsilon_r^2\cA_{\pm}$ 
can be expanded in a power series in $\varepsilon_r$ as 
\begin{eqnarray}
\cA_+^{\prime}&=&\cA_+^{(0)}+\varepsilon_r\cA_+^{(1)}+\varepsilon_r^2\cA_+^{(2)}+\ldots, \label{eq:EI} \\
\cA_-^{\prime}&=&\cA_-^{(0)}+\varepsilon_r\cA_-^{(1)}+\varepsilon_r^2\cA_-^{(2)}+\ldots, \label{eq:EJ}
\end{eqnarray}
then we can try to solve the equations of motion order by order in $\varepsilon_r$. In the $\varepsilon_r\to 0$ 
limit we can expect that the full solutions $\cA_{\pm}^{\prime}$ are well approximated by 
terms of lowest orders. At order $(\varepsilon_r)^0$, we have 
\begin{eqnarray}
\langle\varphi_+|(\Qmid\cA_+^{(0)}+\cA_+^{(0)}*\cA_+^{(0)}-\cA_-^{(0)}*\cA_-^{(0)})\rangle=0, \label{eq:EK}\\
\langle\varphi_-|(\Qmid\cA_-^{(0)}+\cA_+^{(0)}*\cA_-^{(0)}-\cA_-^{(0)}*\cA_+^{(0)})\rangle=0. \label{eq:EL}
\end{eqnarray}
These equations admit a solution $\cA_-^{(0)}=0$ with $\cA_+^{(0)}\neq 0$. Then we are left with 
\begin{equation}
\langle\varphi_+|(\Qmid\cA_+^{(0)}+\cA_+^{(0)}*\cA_+^{(0)})\rangle=0, \label{eq:ELa}
\end{equation}
which has exactly the same form as the equation of motion of bosonic VSFT. It is known that 
this equation is solved by the `$bc$-twisted sliver state'~\cite{GRSZ1} of ghost number 1, 
\begin{eqnarray}
& &|\cA_+^{(0)}\rangle=\cN_+^{(0)}|\Xi^{\prime}\rangle, \label{eq:ELb} \\
& &\langle\Xi^{\prime}|\varphi\rangle\equiv\lim_{n\to\infty}\langle f^{(n)}\circ\varphi^{\prime}(0)
\rangle^{\prime}_{\mathrm{UHP}}=\lim_{n\to\infty}\kappa^{(n)}\langle \tilde{f}^{(n)}\circ\varphi^{\prime}(0)
\rangle^{\prime}_{\mathrm{UHP}}, \label{eq:EM}
\end{eqnarray}
where the conformal maps $f^{(n)},\tilde{f}^{(n)}$ are defined by 
\begin{eqnarray}
f^{(n)}(z)&=&h^{-1}(h(z)^{2/n}), \label{eq:EN} \\
\tilde{f}^{(n)}(z)&=&M_n\circ f^{(n)}(z)=\frac{n}{2}h^{-1}(h(z)^{2/n}),\quad \left( M_n(z)=\frac{n}{2}z\right), \nonumber
\end{eqnarray}
and $|\varphi\rangle=\varphi^{\prime}(0)|0^{\prime}\rangle$ with $|0^{\prime}\rangle$ denoting the $SL(2,\aaru)$-invariant 
vacuum of the twisted conformal field theory CFT$^{\prime}$ and $\langle\ldots\rangle^{\prime}_{\mathrm{UHP}}$ 
represents the correlation function on the upper half plane in CFT$^{\prime}$: See~\cite{GRSZ1} for more detail. 
Two expressions in~(\ref{eq:EM}) agree with each other thanks to the $SL(2,\aaru)$-invariance of the correlation 
function, but we have put a multiplicative factor $\kappa^{(n)}$ which might arise due to the conformal anomaly 
because CFT$^{\prime}$ has a non-vanishing total central charge $c^{\prime}=24$. The advantage of defining the sliver 
as the $n\to\infty$ limit of finite-$n$ wedge states is that we can explicitly compute the $*$-product using the 
generalized gluing and resmoothing theorem, even when some operators are inserted on the sliver. 
The normalization factor $\cN_+^{(0)}$ in front of $\Xi^{\prime}$ 
is determined by the equation of motion~(\ref{eq:ELa}) as 
\begin{equation}
\cN_+^{(0)}=-\frac{1}{e^{2c^{\prime}K}K^{(3)}|f^{(3)\prime}_1(i)|^{1/4}|(h\circ\tilde{f}^{(\infty)})^{\prime}(i)|^{-1/4}
|h\circ\tilde{f}^{(\infty)}(i)-h\circ\tilde{f}^{(\infty)}(-i)|^{1/4}}, \label{eq:EO}
\end{equation}
though its precise expression is not important. A constant $K^{(3)}$ arises when we relate correlation functions 
in the untwisted CFT and those in the twisted CFT$^{\prime}$~\cite{GRSZ1,4138}: 
\begin{eqnarray}
& &\langle f^{(3)}_1\circ\Phi_1(0)f^{(3)}_2\circ\Phi_2(0)f^{(3)}_3\circ\Phi_3(0)\rangle_{\mathrm{UHP}} \label{eq:EP} \\
& &\hspace{5mm} =K^{(3)}\langle f^{(3)}_1\circ\Phi_1^{\prime}(0)f^{(3)}_2\circ\Phi_2^{\prime}(0)
f^{(3)}_3\circ\Phi_3^{\prime}(0)e^{\frac{1}{2}\rho(i)}e^{\frac{1}{2}\bar{\rho}(i)}\rangle_{\mathrm{UHP}}^{\prime}, 
\nonumber 
\end{eqnarray}
where $\rho$ is the bosonized ghost field defined as $c(z)=e^{\rho(z)},b(z)=e^{-\rho(z)}$ with the OPE 
$\rho(z)\rho(0)\sim\log z$, whereas another constant $K$ arises when we use the gluing theorem in a conformal 
field theory with a non-vanishing central charge $c$~\cite{SchSen}: 
\begin{eqnarray}
& &\sum_r\langle f_1\circ\Phi_{r_1}(0)\ldots f_n\circ\Phi_{r_n}(0)f\circ\Phi_r(0)\rangle_{\cD_1}
\langle g_1\circ\Phi_{s_1}(0)\ldots g_m\circ\Phi_{s_m}(0)g\circ\Phi_r^c(0)\rangle_{\cD_2} \phantom{inferno} \nonumber \\
& &\hspace{0mm} =e^{cK}\langle F_1\circ f_1\circ\Phi_{r_1}(0)\ldots F_1\circ f_n\circ\Phi_{r_n}(0)
\widehat{F}_2\circ g_1\circ\Phi_{s_1}(0)\ldots \widehat{F}_2\circ g_m\circ\Phi_{s_m}(0)\rangle_{\cD}.  \label{eq:EQ}
\end{eqnarray}
\medskip

\begin{figure}[htbp]
	\begin{center}
	\scalebox{0.5}[0.5]{\includegraphics{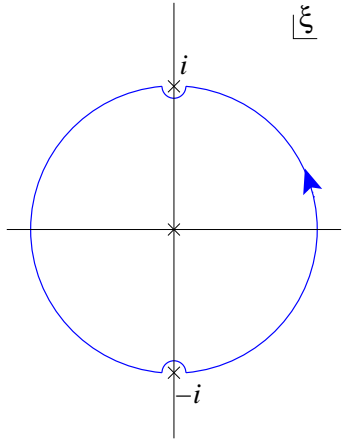}}
	\end{center}
	\caption{The closed contour $\cC$ along the open string $|\xi|=1$.}
	\label{fig:contour}
\end{figure}

\begin{figure}[htbp]
	\begin{center}
	\scalebox{0.6}[0.6]{\includegraphics{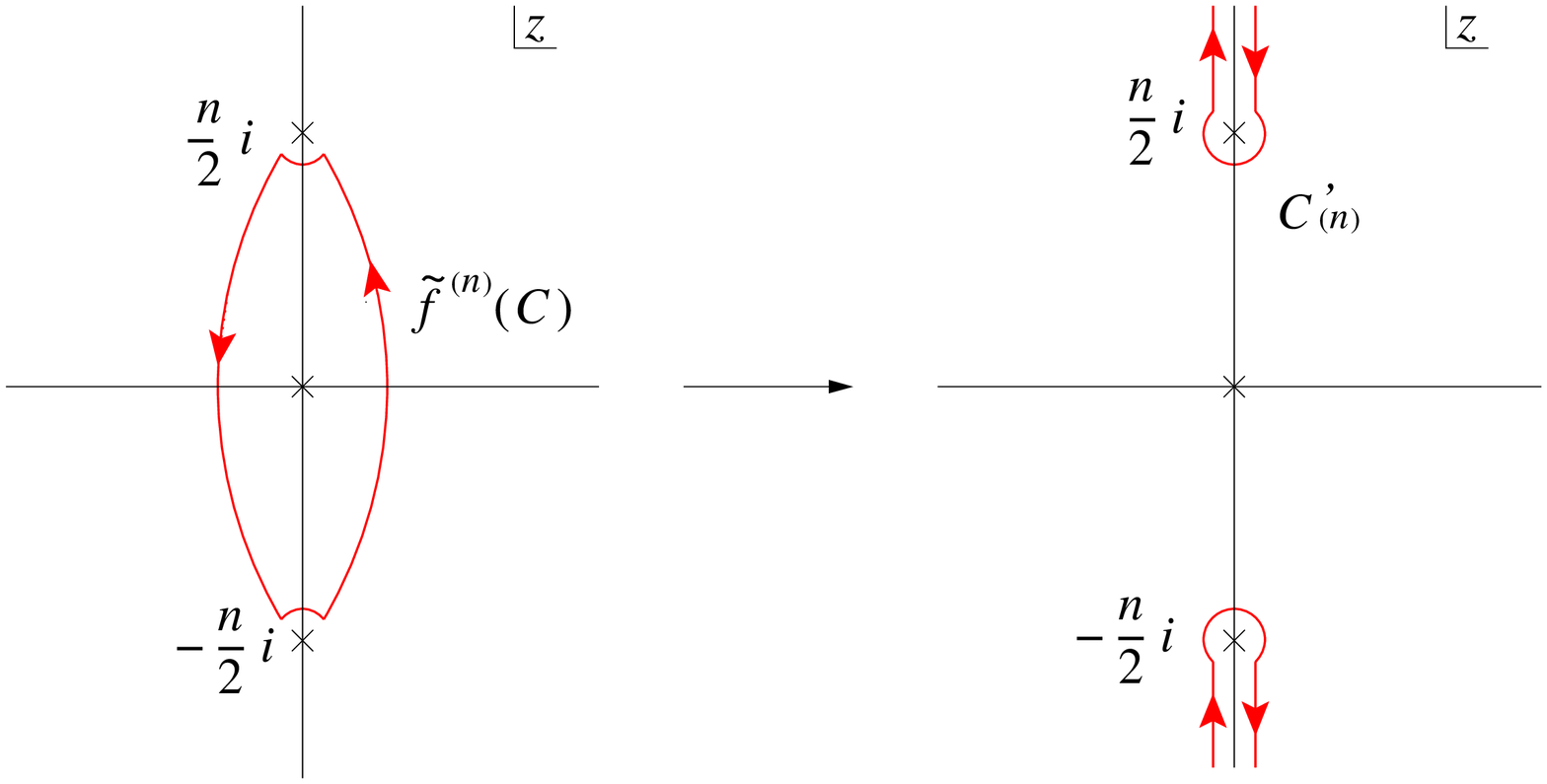}}
	\end{center}
	\caption{Deformation of the integration contour from $\tilde{f}^{(n)}(\cC)$ to $\cC^{\prime}_{(n)}$.}
	\label{fig:deform}
\end{figure}

\begin{figure}[htbp]
	\begin{center}
	\scalebox{0.6}[0.6]{\includegraphics{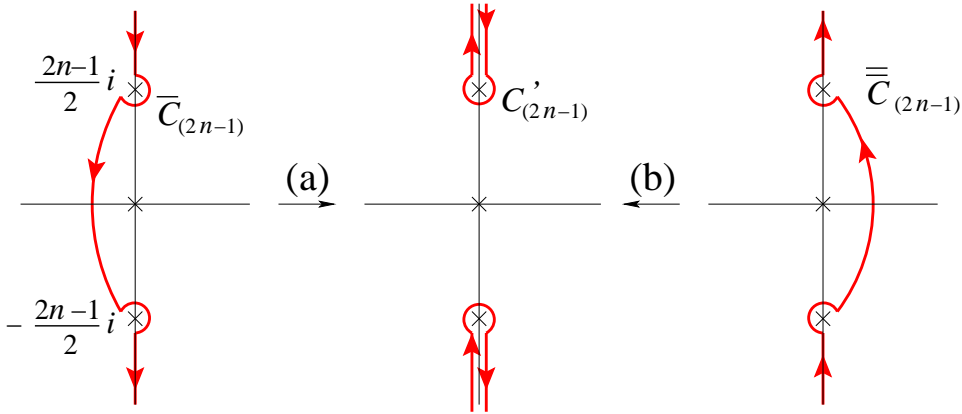}}
	\end{center}
	\caption{We obtain the contours $\overline{\cC}_{(2n-1)},\overline{\overline{\cC}}_{(2n-1)}$ just after 
	applying the gluing theorem. They can be  deformed into $\cC^{\prime}_{(2n-1)}$.}
	\label{fig:deform2}
\end{figure}

At order $(\varepsilon_r)^1$, the equations of motion become 
\begin{eqnarray}
& &\langle\varphi_+|(\Qmid\cA_+^{(1)}+\cA_+^{(0)}*\cA_+^{(1)}+\cA_+^{(1)}*\cA_+^{(0)})\rangle=0, \label{eq:ER} \\
& &\langle\varphi_-|(\Qmid\cA_-^{(1)}+\Gamma\cA_+^{(0)}+\cA_+^{(0)}*\cA_-^{(1)}-\cA_-^{(1)}*\cA_+^{(0)})\rangle=0, 
\label{eq:ES}
\end{eqnarray}
where we have already used the result $\cA_-^{(0)}=0$ at the zeroth order. First we note that eq.(\ref{eq:ER}) is 
solved by $\cA_+^{(1)}=0$. In fact, $(\ref{eq:ELa})+\varepsilon_r\times (\ref{eq:ER})$ gives 
\[ \langle\varphi_+|\left(\Qmid(\cA_+^{(0)}+\varepsilon_r\cA_+^{(1)})+(\cA_+^{(0)}+\varepsilon_r\cA_+^{(1)})*
(\cA_+^{(0)}+\varepsilon_r\cA_+^{(1)})\right)\rangle=0 \]
up to the $\varepsilon_r^2$-term. Since $\cA_+^{(0)}$ and $\cA_+^{(0)}+\varepsilon_r\cA_+^{(1)}$ satisfy 
the same equation, $\cA_+^{(1)}$ should be equal to zero. Next we compute each term in the left hand side of 
eq.(\ref{eq:ES}). We begin by showing that 
\begin{eqnarray}
\langle\varphi_-|\Qmid|\psi_-\rangle&=&\langle\psi_-|\Qmid|\varphi_-\rangle, \label{eq:ET} \\
\langle\varphi_-|\Gamma^{\mathrm{GSO}(+)}|\psi_+\rangle&=&\langle\psi_+|\Gamma^{\mathrm{GSO}(-)}
|\varphi_-\rangle, \label{eq:EU} 
\end{eqnarray}
hold for any Fock space states $\varphi_-$ and $\psi_{\pm}$, with the subscripts $\pm$ denoting their GSO parities. 
The left hand side of~(\ref{eq:ET}) can be handled as 
\begin{eqnarray*}
\langle\varphi_-|\Qmid|\psi_-\rangle&=&\frac{1}{2i}\langle\varphi_-(0)\cR_{\pi}\circ(c(i)-c(-i))\cR_{\pi}\circ\psi_-(0)
\rangle_{\mathrm{UHP}} \\ &=&\frac{1}{2i}\langle\psi_-(0)(\cR_{\pi}\circ\cR_{-\pi})\circ(c(i)-c(-i))
\cR_{-\pi}\circ\varphi_-(0)\rangle_{\mathrm{UHP}} \\ &=&\frac{1}{2i}\langle\psi_-(0)\cR_{\pi}\circ
(-c(i)+c(-i))(-\cR_{\pi}\circ\varphi_-(0))\rangle_{\mathrm{UHP}} \\
&=&\langle\psi_-|\Qmid|\varphi_-\rangle, 
\end{eqnarray*}
where we used the $SL(2,\aaru)$-invariance of the correlator under $\cR_{-\pi}$, and 
the fact that the vertex operators $\varphi_-,\psi_-$ associated with the ghost number 1, GSO($-$) 
states are Grassmann-even. In the same way, 
we find the left hand side of eq.(\ref{eq:EU}) to be 
\begin{eqnarray*}
\langle\varphi_-|\Gamma^{\mathrm{GSO}(+)}|\psi_+\rangle&=&\frac{q_1}{2i}\langle\varphi_-(0)\cR_{\pi}\circ (\gamma(i)-\gamma(-i))
\cR_{\pi}\circ\psi_+(0)\rangle_{\mathrm{UHP}} \\ &=&\frac{q_1}{2i}\langle\psi_+(0)(\cR_{\pi}\circ\cR_{-\pi})\circ
(\gamma(i)-\gamma(-i))\cR_{-\pi}\circ\varphi_-(0)\rangle_{\mathrm{UHP}} \\ &=& \frac{q_1}{2i}\langle\psi_+(0)
\cR_{\pi}\circ ((-i)\gamma(i)-i\gamma(-i))(-\cR_{\pi}\circ\varphi_-(0))\rangle_{\mathrm{UHP}} \\ &=&\langle\psi_+(0)
\cR_{\pi}\circ\left(\frac{q_1}{2}(\gamma(i)+\gamma(-i))\right)\cR_{\pi}\circ\varphi_-(0)\rangle_{\mathrm{UHP}} \\
&=&\langle\psi_+|\Gamma^{\mathrm{GSO}(-)}|\varphi_-\rangle. 
\end{eqnarray*}
Although $\cA_+^{(0)}$ and $\cA_-^{(1)}$ are not Fock space states, we should consider them to satisfy eqs.(\ref{eq:ET}), 
(\ref{eq:EU}) with $\psi_{\pm}$ replaced by $\cA_{\pm}$. Now, we make an ansatz for the solution $\langle\cA_-^{(1)}|$ as 
\begin{equation}
\langle\cA_-^{(1)}|\varphi\rangle=\lim_{n\to\infty}\left\langle\oint\limits_{f^{(n)}(\cC)}\frac{dz}{2\pi i}
s(z)b^{\prime}(z)\gamma(z)\ f^{(n)}\circ\varphi^{\prime}(0)\right\rangle^{\prime}_{\mathrm{UHP}}, \label{eq:EV}
\end{equation}
or conveniently 
\begin{equation}
\langle\cA_-^{(1)}|=\langle\Xi^{\prime}|\oint_{\cC}\frac{d\xi}{2\pi i}s(\xi)b^{\prime}\gamma(\xi). \label{eq:EW}
\end{equation}
Here $s(z)$ is a globally defined holomorphic field of conformal weight $1/2$ such that $dz\ s(z)b^{\prime}\gamma(z)$ 
transforms as a 1-form in CFT$^{\prime}$. $s(z)$ is required to be holomorphic everywhere, except at the puncture $z=0$, 
in the global complex $z$-plane which is obtained as a double-cover of the upper half plane for interacting open strings. 
For $s(z)$ to be regular at infinity, $\lim_{z\to\infty}z s(z)$ must be finite. 
Since $\gamma$ field in this representation is periodic in the NS sector, $s(z)$ must also be periodic under $z\to e^{2\pi i}z$ 
for the integral to be well-defined. Furthermore, $s(z)$ must be a Grassmann-even quantity because the GSO($-$) state 
$\langle\cA_-^{(1)}|$ of ghost number 1 should be Grassmann-even. (Note that $\Xi^{\prime}$ 
itself is Grassmann-odd.) The integration contour $\cC$ goes along the double cover of the `open string' $|\xi|=1$ 
counterclockwise. More precisely, anticipating the insertions at the open string midpoint $z=i$ and its mirror image 
$z=-i$, we define the contour $\cC$ to be the one indicated in Figure~\ref{fig:contour}. 
That is to say, the points $\pm i$ lie \textit{outside} the contour, where we are refering to the left-side 
region of an oriented closed contour, when we walk along it, as the `inside' of the contour. The overall normalization 
of $\cA_-^{(1)}$ has been absorbed in the definition of $s$. At last we compute 
\begin{eqnarray*}
& &\langle\varphi_-|\Qmid|\cA_-^{(1)}\rangle=\langle\cA_-^{(1)}|\Qmid|\varphi_-\rangle \\
& &\hspace{3mm} =\lim_{n\to\infty}\kappa^{(n)}\left\langle\oint\limits_{\tilde{f}^{(n)}(\cC)}
\frac{dz}{2\pi i}s(z)b^{\prime}\gamma(z)\frac{1}{2}\left(c^{\prime}(\tilde{f}^{(n)}(i))+c^{\prime}
(\tilde{f}^{(n)}(-i))\right)\tilde{f}^{(n)}\circ\varphi_-^{\prime}(0)\right\rangle^{\prime}_{\mathrm{UHP}}, 
\end{eqnarray*}
where we have used eqs.(\ref{eq:ET}), (\ref{eq:EV}) and (\ref{eq:EM}). In CFT$^{\prime}$ $\Qmid$ is written 
as $\frac{1}{2}(c^{\prime}(i)+c^{\prime}(-i))$ and $c^{\prime}$ has conformal weight 0. Since $sb^{\prime}\gamma(z)$ 
is holomorphic everywhere except at the origin, the integration contour $\tilde{f}^{(n)}(\cC)$ can be deformed 
in the way indicated in Figure~\ref{fig:deform}. 
We call the new contour $\cC^{\prime}_{(n)}$. In the limit $n\to\infty$, both $\tilde{f}^{(n)}(i)=ni/2$ and 
$\tilde{f}^{(n)}(-i)=-ni/2$ go to the same point at infinity and $\cC^{\prime}_{(\infty)}$ becomes a small 
contour encircling the infinity in the clockwise direction. Picking up the pole coming from the operator product 
of $b^{\prime}$ with $c^{\prime}$, we obtain 
\begin{equation}
\langle\varphi_-|\Qmid|\cA_-^{(1)}\rangle=-\kappa^{(\infty)}s(\infty)\langle\gamma(\infty)
\tilde{f}^{(\infty)}\circ\varphi_-^{\prime}(0)\rangle^{\prime}_{\mathrm{UHP}}. \label{eq:EX}
\end{equation}
The next one to calculate is 
\[\langle\varphi_-|\Gamma|\cA_+^{(0)}\rangle =\langle\cA_+^{(0)}|\Gamma^{\mathrm{GSO}(-)}|\varphi_-\rangle 
=\frac{q_1}{2}\cN_+^{(0)}\langle\Xi^{\prime}|(\gamma(i)+\gamma(-i))|\varphi_-\rangle.\]
This is easily evaluated to give 
\begin{equation}
\langle\varphi_-|\Gamma|\cA_+^{(0)}\rangle =\frac{q_1}{2}\cN_+^{(0)}\kappa^{(\infty)}\left(
[\tilde{f}^{(\infty)\prime}(i)]^{-\frac{1}{2}}+[\tilde{f}^{(\infty)\prime}(-i)]^{-\frac{1}{2}}\right)
\langle\gamma(\infty)\tilde{f}^{(\infty)}\circ\varphi_-^{\prime}(0)\rangle^{\prime}_{\mathrm{UHP}}. \label{eq:EY}
\end{equation}
We then turn to the third term in~(\ref{eq:ES}). Inserting the complete set of states \linebreak
$\mathbf{1}=\sum_r|\Phi_r\rangle\langle\Phi_r^c|$, we have 
\begin{eqnarray}
& &\langle\varphi_-|\cA_+^{(0)}*\cA_-^{(1)}\rangle=\cN_+^{(0)}\sum_{r,s}\langle\varphi_-|\left(
|\Phi_r\rangle*\left(\cR_{-\pi}\circ\oint_{\cC}\frac{d\xi}{2\pi i}s(\xi)b^{\prime}\gamma(\xi)\right)
|\Phi_s\rangle\right) \langle\Phi_r^c|\Xi^{\prime}\rangle \langle\Phi_s^c|\Xi^{\prime}\rangle \nonumber \\
& &\hspace{1cm}=\cN_+^{(0)}\sum_{r,s}\left\langle f^{(3)}_1\circ\varphi_-(0)f^{(3)}_2\circ\Phi_r(0)f^{(3)}_3\circ
\left[\cR_{-\pi}\circ\left(\oint_{\cC}\frac{d\xi}{2\pi i}s(\xi)\xi b(\xi)\gamma(\xi)\right)\ \Phi_s(0)\right]
\right\rangle_{\mathrm{UHP}} \nonumber \\
& &\hspace{3cm}\times\lim_{n\to\infty}\langle f^{(n)}\circ\Phi_r^{c\prime}(0)\rangle^{\prime}_{\mathrm{UHP}}
\langle f^{(n)}\circ\Phi_s^{c\prime}(0)\rangle^{\prime}_{\mathrm{UHP}}, \label{eq:EZ}
\end{eqnarray}
where we have used 
\[ |\cA_-^{(1)}\rangle=\cR_{-\pi}\circ\left(\oint_{\cC}\frac{d\xi}{2\pi i}s(\xi)b^{\prime}\gamma(\xi)\right)
|\Xi^{\prime}\rangle. \]
After rewriting the CFT correlator in~(\ref{eq:EZ}) as a CFT$^{\prime}$ correlator using eq.(\ref{eq:EP}), 
we can express (\ref{eq:EZ}) as a single correlator by making use of the gluing theorem~(\ref{eq:EQ}) 
twice. From the formulas given in~\cite{KisOhm} we obtain the result 
\begin{eqnarray}
& &\langle\varphi_-|\cA_+^{(0)}*\cA_-^{(1)}\rangle=\cN_+^{(0)}K^{(3)}|f^{(3)\prime}_1(i)|^{1/4}e^{2c^{\prime}K}
\label{eq:FA} \\ & &\hspace{1.5cm} \times\lim_{n\to\infty}\kappa^{(n)}\left\langle\oint_{\cC^{\prime}_{(2n-1)}}
\frac{dz}{2\pi i}
s(z)b^{\prime}\gamma(z)\ \tilde{f}^{(2n-1)}\circ\left( e^{\frac{1}{2}\rho(i)}e^{\frac{1}{2}\rho(-i)}
\varphi_-^{\prime}(0)\right)\right\rangle^{\prime}_{\mathrm{UHP}}, \nonumber
\end{eqnarray}
where the integration contour, after the deformation shown in Figure~\ref{fig:deform2}(a), has become 
the one $\cC^{\prime}_{(2n-1)}$ previously defined. 
In the $n\to\infty$ limit $e^{\frac{1}{2}\rho(\tilde{f}^{(2n-1)}(i))}$ and $e^{\frac{1}{2}\rho(\tilde{f}^{(2n-1)}(-i))}$ 
come close to each other so that their product can be replaced by the leading term $e^{\rho}=c^{\prime}$ in the OPE. 
This fact can be made more transparent by moving from the upper half plane to the unit disk representation 
by the conformal map $h(z)$. Performing the contour integration, we finally reach 
\begin{eqnarray}
\langle\varphi_-|\cA_+^{(0)}*\cA_-^{(1)}\rangle&=&-\cN_+^{(0)}K^{(3)}e^{2c^{\prime}K}\kappa^{(\infty)}
|f^{(3)\prime}_1(i)|^{1/4}|(h\circ\tilde{f}^{(\infty)})^{\prime}(i)|^{-1/4} \label{eq:FB} \\
& &\times |h\circ\tilde{f}^{(\infty)}(i)-h\circ\tilde{f}^{(\infty)}(-i)|^{1/4}s(\infty)\langle
\gamma(\infty)\tilde{f}^{(\infty)}\circ\varphi_-^{\prime}(0)\rangle^{\prime}_{\mathrm{UHP}}. \nonumber 
\end{eqnarray}
The last term in eq.(\ref{eq:ES}) is found to be 
\begin{eqnarray}
& &-\langle\varphi_-|\cA_-^{(1)}*\cA_+^{(0)}\rangle=-\cN_+^{(0)}\sum_{r,s}\lim_{n\to\infty}
\langle f^{(n)}\circ\Phi_r^{c\prime}(0)\rangle^{\prime}_{\mathrm{UHP}}
\langle f^{(n)}\circ\Phi_s^{c\prime}(0)\rangle^{\prime}_{\mathrm{UHP}} \label{eq:FC} \\
& &\hspace{1cm} \times\left\langle f^{(3)}_1\circ\varphi_-(0)f^{(3)}_2\circ
\left(\cR_{-\pi}\circ\oint_{\cC}\frac{d\xi}{2\pi i}s(\xi)\xi b(\xi)\gamma(\xi)\right)
f^{(3)}_2\circ\Phi_r(0)f^{(3)}_3\circ\Phi_s(0)\right\rangle_{\mathrm{UHP}}. \nonumber 
\end{eqnarray}
For $|\Phi_r\rangle\langle\Phi_r^c|\Xi^{\prime}\rangle$ to be non-vanishing, $\Phi_r$ must 
share its properties with $\Xi^{\prime}$. In particular, $\Phi_r(0)$ is Grassmann-odd. 
So when $f^{(3)}_2\circ\Phi_r(0)$ passes through 
$f^{(3)}_2\circ\cR_{-\pi}\circ\oint_{\cC}\frac{d\xi}{2\pi i}s(\xi)\xi b(\xi)\gamma(\xi)$ 
there arises a minus sign, which cancels the overall minus sign in front of the right hand side of~(\ref{eq:FC}). 
Comparing it with eq.(\ref{eq:EZ}), we find that $\langle\varphi_-|\cA_+^{(0)}*\cA_-^{(1)}\rangle$ and 
$-\langle\varphi_-|\cA_-^{(1)}*\cA_+^{(0)}\rangle$ give the same expression except for the integration contour. 
Calculating it further, we get the same expression as eq.(\ref{eq:FA}), but with the contour replaced by 
$\overline{\overline{\cC}}_{(2n-1)}$ shown in Figure~\ref{fig:deform2}. This contour, however, can be deformed 
into $\cC^{\prime}_{(2n-1)}$ (Figure~\ref{fig:deform2}(b)) due to the holomorphicity, so 
$-\langle\varphi_-|\cA_-^{(1)}*\cA_+^{(0)}\rangle$ gives exactly the same result as 
$\langle\varphi_-|\cA_+^{(0)}*\cA_-^{(1)}\rangle$ (\ref{eq:FB}). Substituting our results~(\ref{eq:EX}), 
(\ref{eq:EY}), (\ref{eq:FB}) into the equation of motion~(\ref{eq:ES}), we find that it is solved if we choose 
\begin{equation}
s(\infty)\equiv s\left(z=\tilde{f}^{(\infty)}(i)=\infty\right)=-\frac{q_1}{2}\cN_+^{(0)}\left( 
[\tilde{f}^{(\infty)\prime}(i)]^{-1/2}+[\tilde{f}^{(\infty)\prime}(-i)]^{-1/2}\right), \label{eq:FD}
\end{equation}
where we have used the explicit form~(\ref{eq:EO}) of $\cN_+^{(0)}$ to simplify the expression. Again, what 
is important is not the precise expression~(\ref{eq:FD}) for $s(\infty)$ but the fact that we could solve 
the equations of motion~(\ref{eq:ER}), (\ref{eq:ES}) at order $(\varepsilon_r)^1$ by $\cA_+^{(1)}=0$ and 
$\cA_-^{(1)}$ given in~(\ref{eq:EV}) with a suitable choice of a parameter $s(\infty)$. 
\medskip 

Since the above $s(\infty)$ looks like a constant, one might suspect that this solution does not 
satisfy the regularity at infinity: $\lim_{z\to\infty}zs(z)<\infty$. However, 
$\tilde{f}^{(\infty)\prime}(z)=1/(1+z^2)$ diverges when $z\to \pm i$ so that $s(\infty)$ is in fact 
vanishingly small. Regularizing the expression by $i\to i+\epsilon$ and taking the contribution from 
$\cN_+^{(0)}$ into account, we find 
\begin{eqnarray*}
\lim_{z\to\infty}zs(z)&\equiv&\lim_{\epsilon\to 0}\tilde{f}^{(\infty)}(i+\epsilon)s(\tilde{f}^{(\infty)}(i+\epsilon)) \\
&\sim&\lim_{\epsilon\to 0}\log\epsilon\left(\epsilon^{\frac{1}{12}-\frac{1}{4}}(\log\epsilon)^{\frac{1}{4}}\right)
\epsilon^{\frac{1}{2}}\sim\lim_{\epsilon\to 0}\epsilon^{\frac{1}{3}}=0, 
\end{eqnarray*}
where we have used $\tilde{f}^{(\infty)}(i+\epsilon)=\tan^{-1}(i+\epsilon)\simeq\frac{1}{2i}\log\frac{i\epsilon}{2}
\sim\log\epsilon, \tilde{f}^{(n)\prime}(i+\epsilon)\sim\epsilon^{\frac{2}{n}-1}$. 
Therefore the choice~(\ref{eq:FD}) of $s(\infty)$ does not violate the regularity condition. On the other hand, 
the equations of motion at order $(\varepsilon_r)^2$ have not constrained the functional form of $s(z)$ except 
its value at the open string midpoint, as long as the chosen $s(z)$ satisfies the requirements mentioned 
below eq.(\ref{eq:EW}). We cannot tell without solving the equations of motion at higher orders to what degree 
the form of $s$ should be determined. 
\smallskip

Our approximate solution 
$\widehat{\cA}^{\prime}=\cA_+^{(0)}\otimes\sigma_3+\varepsilon_r\cA_-^{(1)}\otimes i\sigma_2$ 
given in (\ref{eq:ELb}), (\ref{eq:EV}) has the following properties: 
\begin{itemize}
  \item At the lowest order $(\varepsilon_r)^0$ the solution $\cA_+^{(0)}$ takes the same form as the twisted 
  sliver solution of bosonic vacuum string field theory, which is supposed to represent a spacetime-filling 
  D25-brane;
  \item Both $\cA_+^{(0)}$ and $\cA_-^{(1)}$ have the matter-ghost split form and their matter part is 
  simply the matter sliver~\cite{KP,RSZ2,RSZ4,NSms,MS}.
\end{itemize}
These properties suggest that, if we can extend it to all orders in $\varepsilon_r$, the full solution 
should represent a spacetime-filling non-BPS D9-brane of type IIA superstring theory.\footnote{We are 
assuming that we take the boundary conformal field theory associated with a non-BPS D9-brane of type IIA 
theory as the reference BCFT in whose state space the classical string field takes value.} 

\sectiono{Ghost Kinetic Operator and Solutions in Non-polynomial Vacuum Superstring Field 
Theory}\label{sec:nonpoly}
In this section we consider Berkovits' non-polynomial superstring field theory around the tachyon vacuum. 
First of all, let us note that, if we expand the GSO-unprojected string field $\widehat{\Phi}$ 
as $e^{\widehat{\Phi}}=e^{\widehat{\Phi}_0}*e^{\widehat{\phi}}$ around a classical solution 
$\widehat{\Phi}_0$ of the equation of motion $\widehat{\eta}_0(e^{-\widehat{\Phi}}\widehat{Q}_Be^{\widehat{\Phi}})=0$ 
derived from the Wess-Zumino-Witten--like action~\cite{sP,BSZ}
\begin{eqnarray}
\hspace{-2cm}
S&=&\frac{1}{4g_o^2}\mathrm{Tr}\bllk\left(e^{-\widehat{\Phi}}\widehat{Q}_B
e^{\widehat{\Phi}}\right)\left(e^{-\widehat{\Phi}}\widehat{\eta}_0
e^{\widehat{\Phi}}\right) \nonumber \\
& &{}-\int_0^1dt\left(e^{-t\widehat{\Phi}}\partial_te^{t\widehat{\Phi}}\right)
\left\{\left(e^{-t\widehat{\Phi}}\widehat{Q}_Be^{t\widehat{\Phi}}\right),
\left(e^{-t\widehat{\Phi}}\widehat{\eta}_0e^{t\widehat{\Phi}}\right)\right\}
\brrk,  \label{eq:FG}
\end{eqnarray}
then the action for the fluctuation field $\widehat{\phi}$ takes the same form as the original one~(\ref{eq:FG}) 
with the BRST operator $\widehat{Q}_B$ replaced by another operator $\widehat{Q}$ given by 
\begin{equation}
\widehat{Q}\widehat{X}=\widehat{Q}_B\widehat{X}+\widehat{A}_0*\widehat{X}-(-1)^{\gh (\widehat{X})}
\widehat{X}*\widehat{A}_0, \label{eq:FH}
\end{equation}
where we have defined $\widehat{A}_0=e^{-\widehat{\Phi}_0}\widehat{Q}_Be^{\widehat{\Phi}_0}$. 
This is a trivial extension of the results shown in~\cite{Klu1,MS} for the GSO-projected theory. 
Since the above formula~(\ref{eq:FH}) for 
$\widehat{Q}$ is very similar to the one~(\ref{eq:AK}) for the cubic case and both cubic and Berkovits' 
superstring field theories reproduce qualitatively the same tachyon potential of the double-well 
form~\cite{0011117,BSZ}, the logic we presented in section~\ref{sec:proposal} should be valid in this 
non-polynomial case as well: We need in $\widehat{Q}$ the Grassmann-even sector $\Qe\otimes i\sigma_2$ 
as well as the Grassmann-odd piece $\Qo\otimes\sigma_3$ 
to violate the $\zetto_2$ reflection symmetry, and through a singular reparametrization $\widehat{\cQ}$ 
should be dominated by the midpoint insertions of the lowest-dimensional local operators. 
Moreover, Berkovits' superstring field theory action~(\ref{eq:FG}) including both the GSO($\pm$) sectors 
was shown to be invariant under the twist operation~\cite{BSZ} 
\begin{equation}
\Omega  |\Phi\rangle =\left\{
  \begin{array}{ll}
       (-1)^{h_{\Phi}+1}|\Phi\rangle  &  \mbox{for GSO($+$) states } (h_{\Phi}\in \zetto)  \\
      (-1)^{h_{\Phi}+\frac{1}{2}}|\Phi\rangle  &  \mbox{for GSO($-$) states } (h_{\Phi}\in\zetto+\frac{1}{2}).  
  \end{array}
\right., \label{eq:FK}
\end{equation}
which is the same as the action of the twist operator~(\ref{eq:AZ}) in the 0-picture cubic superstring 
field theory. Hence, if we require the kinetic operator $\widehat{\cQ}$ around the tachyon vacuum to commute 
with the twist operator $\Omega$, we will obtain the same $\widehat{\cQ}$ as in the cubic case and 
the resulting action will become twist-invariant under~(\ref{eq:FK}). From these considerations, 
we propose that the non-polynomial vacuum superstring field theory action for the NS sector is given by 
\begin{eqnarray}
\hspace{-2cm}
S&=&\frac{\kappa_0}{4}\mathrm{Tr}\bllk\left(e^{-\widehat{\Phi}}\widehat{\cQ}
e^{\widehat{\Phi}}\right)\left(e^{-\widehat{\Phi}}\widehat{\eta}_0
e^{\widehat{\Phi}}\right) \nonumber \\
& &{}-\int_0^1dt\left(e^{-t\widehat{\Phi}}\partial_te^{t\widehat{\Phi}}\right)
\left\{\left(e^{-t\widehat{\Phi}}\widehat{\cQ}e^{t\widehat{\Phi}}\right),
\left(e^{-t\widehat{\Phi}}\widehat{\eta}_0e^{t\widehat{\Phi}}\right)\right\}
\brrk,  \label{eq:FL}
\end{eqnarray}
with
\begin{eqnarray}
& &\widehat{\cQ}=\Qod\otimes\sigma_3-\Qev\otimes i\sigma_2, \nonumber \\
& &\hspace{3mm} \Qod=\frac{1}{2i\varepsilon_r^2}(c(i)-c(-i))+\frac{q_1^2}{2}\oint\frac{dz}{2\pi i}
b\gamma^2(z), \label{eq:FM} \\
& &\hspace{3mm} \Qev^{\mathrm{GSO}(+)}=\frac{q_1}{2i\varepsilon_r}(\gamma(i)-\gamma(-i)), \qquad 
\Qev^{\mathrm{GSO}(-)}=\frac{q_1}{2\varepsilon_r}(\gamma(i)+\gamma(-i)). \nonumber 
\end{eqnarray}
This $\widehat{\cQ}$ is a nilpotent derivation of the $*$-algebra, and can be regularized in such a way 
that $\widehat{\cQ}$ annihilates the identity string field $\cI$, as has been proven in subsection~\ref{subsec:Qform}. 
In addition, $\widehat{\cQ}$ anticommutes with $\widehat{\eta}_0=\eta_0\otimes\sigma_3$ because $\widehat{\cQ}$ 
contains no factor of $\xi_0$ and $\widehat{\eta}_0$ does not change the GSO parity of the states. 
These properties guarantee that the action~(\ref{eq:FL}) is invariant under the infinitesimal gauge transformation 
\begin{equation}
\delta (e^{\widehat{\Phi}})=(\widehat{\cQ}\widehat{\Omega})*e^{\widehat{\Phi}}+e^{\widehat{\Phi}}*(\widehat{\eta}_0
\widehat{\Lambda}). \label{eq:FN}
\end{equation}
where the gauge parameters $\widehat{\Omega},\widehat{\Lambda}$ are of ghost number $-1$. Actually we need not 
repeat the proof of gauge invariance here because the set $(\widehat{\Phi},\widehat{\cQ},\widehat{\eta}_0)$ 
of the GSO-unprojected string field $\widehat{\Phi}=\Phi_+\otimes\mathbf{1}+\Phi_-\otimes\sigma_1$ and the operators 
$\widehat{\cQ},\widehat{\eta}_0$ equipped with the internal Chan-Paton 
matrices satisfies exactly the same algebraic relations as that $(\Phi,Q_B,\eta_0)$ in the GSO-projected theory,
so that the proof goes in the identical way to the one given in the GSO-projected theory 
(see~\cite{BSZ}), as can be seen from the processes of the proof presented in subsection~\ref{subsec:gauge}.
\medskip

What we want to do next is to find classical solutions\footnote{Recently, a general method of constructing 
exact solutions of the equation of motion of Berkovits' string field theory was given in~\cite{LPU}.}
corresponding to D-branes. For this purpose, 
we will follow the strategy proposed in~\cite{4138}. Let us consider the specific combination 
$\widehat{A}(\widehat{\Phi})\equiv e^{-\widehat{\Phi}}\widehat{\cQ}e^{\widehat{\Phi}}$ of the 
string field $\widehat{\Phi}$ of this theory. Since $\widehat{\Phi}$ has vanishing ghost and picture numbers, 
we find that 
\begin{equation}
\widehat{A}\mbox{ has ghost number 1 and picture number 0.} \label{eq:FO}
\end{equation}
Written in terms of $\widehat{A}(\widehat{\Phi})$, the equation of motion following from the action~(\ref{eq:FL}) is
\begin{equation}
\widehat{\eta}_0(\widehat{A})=0. \label{eq:FP}
\end{equation}
In addition, $\widehat{A}(\widehat{\Phi})$ satisfies by definition 
\begin{equation}
\widehat{\cQ}\widehat{A}+\widehat{A}*\widehat{A}=0, \label{eq:FQ}
\end{equation}
because $\widehat{\cQ}$ is a nilpotent derivation and annihilates 
$\cI\otimes\mathbf{1}=e^{-\widehat{\Phi}}*e^{\widehat{\Phi}}$. On the other hand, in 0-picture 
cubic superstring field theory the \textit{string field} $\widehat{\cA}$ was defined to have 
\begin{equation}
\mbox{ghost number 1 and picture number 0.}\label{eq:FR}
\end{equation}
Since this cubic theory was formulated within the ``small" Hilbert space, $\widehat{\cA}$ must 
not contain the zero mode of $\xi$. In other words, we have to impose 
\begin{equation}
\widehat{\eta}_0(\widehat{\cA})=0. \label{eq:FS}
\end{equation}
As given in (\ref{eq:EB}), the equation of motion of the cubic theory was 
\begin{equation}
\widehat{\cQ}\widehat{\cA}+\widehat{\cA}*\widehat{\cA}=0. \label{eq:FT}
\end{equation}
Comparing eqs.(\ref{eq:FO})--(\ref{eq:FQ}) with eqs.(\ref{eq:FR})--(\ref{eq:FT}), we find that 
the sets of equations we should solve in looking for classical solutions in these two theories 
coincide with each other.\footnote{Since the picture changing operations have been completely eliminated 
from the theory in Berkovits' formulation, the equation~(\ref{eq:FQ}) should be considered by taking 
the inner product with a state of ghost number 0, picture number $-1$ \textit{including} $\xi_0$, 
with no picture changing operators inserted. Since we considered eq.(\ref{eq:EE}) in section~\ref{sec:solution} 
instead of $\llk\widehat{Y}_{-2}|\widehat{\varphi},\cF(\widehat{\cA})\rrk=0$, the solution found there 
can be used here.}
This fact can be used to find a solution $\widehat{\Phi}_0$ in the non-polynomial theory which `corresponds' to 
a solution $\widehat{\cA}_0$ in the cubic theory. Suppose that we are given a solution $\widehat{\cA}_0$ 
in the cubic theory satisfying~(\ref{eq:FR})--(\ref{eq:FT}). From~(\ref{eq:FR}) and (\ref{eq:FT}), it is 
\textit{possible} that there exists a string field configuration $\widehat{\Phi}_0$ which satisfies 
\begin{equation}
e^{-\widehat{\Phi}_0}\widehat{\cQ}e^{\widehat{\Phi}_0}=\widehat{\cA}_0. \label{eq:FU}
\end{equation}
\textit{If we can find such a} $\widehat{\Phi}_0$, it then gives a solution in the non-polynomial theory 
because we have 
\[ \widehat{\eta}_0(e^{-\widehat{\Phi}_0}\widehat{\cQ}e^{\widehat{\Phi}_0})=\widehat{\eta}_0
(\widehat{\cA}_0)=0, \]
where (\ref{eq:FS}) was used. Furthermore, we expect that the solution $\widehat{\Phi}_0$ 
in the non-polynomial theory has the same physical interpretation as the corresponding one $\widehat{\cA}_0$ 
in the cubic theory. In particular, the forms of the new kinetic operators around these solutions are, 
respectively, 
\[ \widehat{\cQ}^{\prime}_{\mathrm{cubic}}\widehat{X}=\widehat{\cQ}\widehat{X}+\widehat{\cA}_0*\widehat{X}
-(-1)^{\gh(\widehat{X})}\widehat{X}*\widehat{\cA}_0 \quad \mbox{ in cubic theory} \]
and 
\[ \widehat{\cQ}^{\prime}_{\mbox{\scriptsize{non-poly}}}\widehat{X}=\widehat{\cQ}\widehat{X}+
\widehat{A}(\widehat{\Phi}_0)*\widehat{X}
-(-1)^{\gh(\widehat{X})}\widehat{X}*\widehat{A}(\widehat{\Phi}_0) \quad \mbox{ in non-polynomial theory}. \]
Since we have determined $\widehat{\Phi}_0$ such that $\widehat{\cA}_0=\widehat{A}(\widehat{\Phi}_0)$ 
is satisfied, these kinetic operators 
$\widehat{\cQ}^{\prime}_{\mbox{\scriptsize{cubic}}},\widehat{\cQ}^{\prime}_{\mbox{\scriptsize{non-poly}}}$, 
each of which describes the perturbative physics around the corresponding solution, should agree with each other. 
This result strongly supports the claim that the solutions $\widehat{\cA}_0$ and $\widehat{\Phi}_0$ 
of the two theories related through~(\ref{eq:FU}) share a common physical interpretation. 
Therefore, in order to find D-brane solutions $\widehat{\Phi}_0$ in non-polynomial vacuum superstring 
field theory, we have to 
\begin{itemize}
  \item[(1)] find solutions $\widehat{\cA}_0$ in 0-picture cubic vacuum superstring field theory which can be 
  interpreted as D-branes, and then
  \item[(2)] solve the equation~(\ref{eq:FU}) with respect to $\widehat{\Phi}_0$ for $\widehat{\cA}_0$ just found above. 
\end{itemize}
However, we have not yet found any explicit algorithm to solve the equation~(\ref{eq:FU}) for a given 
$\widehat{\cA}_0$. This problem is left to future study. 
\smallskip

Finally we remark that, even if $\widehat{\cA}_0$ can be written in the form 
$e^{-\widehat{\Phi}_0}\widehat{\cQ}e^{\widehat{\Phi}_0}$, it does not immediately follow that 
this $\widehat{\cA}_0$ is pure-gauge in the cubic theory. This is because in Berkovits' formulation 
we can seek an appropriate configuration $\widehat{\Phi}_0$ in the ``large" Hilbert space including $\xi_0$, 
whereas to make an assertion that $\widehat{\cA}_0$ is pure-gauge in the cubic theory we must find 
a suitable gauge parameter $\widehat{\Lambda}$ satisfying 
$\widehat{\cA}_0=e^{-\widehat{\Lambda}}\widehat{\cQ}e^{\widehat{\Lambda}}$ within the ``small" Hilbert space. 
Conversely, let us suppose that we have a pure-gauge configuration $\widehat{\cA}_0$ in the cubic theory. 
Then there exists a gauge parameter $\widehat{\Lambda}$ which has ghost number 0 and satisfies 
$\widehat{\cA}_0=e^{-\widehat{\Lambda}}\widehat{\cQ}e^{\widehat{\Lambda}}$ and 
$\widehat{\eta}_0\widehat{\Lambda}=0$. If we regard this $\widehat{\Lambda}$ as a string field in Berkovits' 
non-polynomial theory, such a configuration (\textit{i.e.} annihilated by $\widehat{\eta}_0$) turns out 
to be pure-gauge, as shown in~\cite{4138}. So the story is quite consistent in the sense that a pure-gauge 
configuration in one theory is mapped under~(\ref{eq:FU}) to some pure-gauge configuration 
in the other theory.\footnote{More generally, we can show that two gauge-equivalent string field 
configurations $e^{\widehat{\Phi}_0},h_1^{\widehat{\cQ}}e^{\widehat{\Phi}_0}h_2^{\widehat{\eta}_0}$ in 
non-polynomial theory are mapped to two configurations 
$\widehat{\cA}_0,(h_2^{\widehat{\eta}_0})^{-1}(\widehat{\cQ}+\widehat{\cA}_0)h_2^{\widehat{\eta}_0}$ 
which are also related through a formally valid gauge transformation in the cubic theory, where 
$h_1^{\widehat{\cQ}},h_2^{\widehat{\eta}_0}$ are gauge parameters annihilated by $\widehat{\cQ},\widehat{\eta}_0$, 
respectively.}

\sectiono{Summary and Discussion}\label{sec:sumdis}
In this paper we have discussed the construction of vacuum superstring field theory itself and the 
brane solutions in it. We have first argued what form the kinetic operator $\widehat{Q}$ around the 
tachyon vacuum should have in general, and seen that we need a Grassmann-even operator $\Qe\otimes i\sigma_2$ 
to have a structure expected of the vacuum superstring field theory action. We have then determined the 
form of the simple pure-ghost kinetic operator $\widehat{\cQ}$ as unambiguously as possible by requiring that 
(i) the most singular parts of $\widehat{\cQ}$ be made up of midpoint insertions of the lowest-dimensional 
operators, following~\cite{GRSZ1}, (ii) $\widehat{\cQ}$ preserve the twist invariance of the string field 
theory action, and (iii) the action be gauge invariant. We have obtained a somewhat surprising result that 
the form of $\Qev$ depends on the GSO parity of the state on which $\Qev$ acts, but such a structure is 
indispensable for consistency. We have examined in detail the properties of $\widehat{\cQ}$ determined this way, 
and explicitly shown that it satisfies all of the axioms which guarantee the gauge invariance of the action. 
We have observed from the known properties of the twist operation that $\widehat{\cQ}$ should take the 
same form in both cubic and non-polynomial types of vacuum superstring field theory. 

We have also tried to solve the equations of motion of cubic vacuum superstring field theory, but 
obtained only an approximate solution in an $\varepsilon_r$-expansion method. Finally we have suggested 
a way of generating the solution in the non-polynomial theory which physically corresponds to 
a given solution in the cubic theory, with its precise algorithm left to be clarified. 
\medskip

To completely understand the nature of superstring field theory around the tachyon vacuum, 
many things still remain to be done. At the level of the determination of the kinetic operator 
$\widehat{\cQ}$, aside from a proportionality constant we have not been able to uniquely fix 
the form of the non-leading part of $\Qod$, which has been added \textit{by hand} in order to 
make $\widehat{\cQ}$ be nilpotent. It might be that, after successfully including the Ramond sector field 
in vacuum superstring field theory, the requirement of spacetime supersymmetry restored around the 
`type II closed string vacuum' puts further constraints on the form of $\widehat{\cQ}$. 

At the level of the construction of classical solutions, we have not obtained exact expressions even for 
the solution representing the most fundamental non-BPS D9-brane. Moreover, the expressions for our approximate 
solution are plagued with many vanishing or diverging factors, as is often the case with the singular 
representative of vacuum string field theory. In the bosonic case, such a `formal' argument was partly 
justified by showing~\cite{GRSZ1,Okuda} that the twisted sliver solution coincides with the Siegel gauge solution found 
in~\cite{HK} in a less singular algebraic approach. So it is desirable that 
we could find a solution algebraically in this superstring case as well and compare it with ours. 
This might also give us insight into the construction of another tachyon vacuum or BPS D-brane solutions.

\section*{Acknowledgements}
I am grateful to I. Kishimoto for many valuable discussions and comments. 
I'd like to thank T. Eguchi and Y. Matsuo for discussions, and A.A. Giryavets for E-mail correspondence. 
I also thank the members of Hokkaido University and the organizers of 
YITP workshop YITP-W-02-04 
on ``Quantum Field Theory 2002", 
where I was given opportunities to have discussions on this research. 
This work is supported by JSPS Research Fellowships for Young Scientists. 


\section*{Appendices}
\renewcommand{\thesection}{\Alph{section}}
\setcounter{section}{0}

\sectiono{Inner Derivation Formula for $\Qev$}\label{sec:appA}
In this Appendix we show that the expression~(\ref{eq:CH}) reproduces eq.(\ref{eq:CI}). Taking the 
inner product of~(\ref{eq:CH}) with a Fock space state $\langle\varphi|$ gives 
\begin{eqnarray}
& &\langle\varphi|\Qev |\psi\rangle=\langle\varphi|(\Gamma_{\epsilon}\cI)*\psi\rangle-(-1)^{\mathrm{GSO}(\psi)}
\langle\varphi|\psi *(\Gamma_{\epsilon}\cI)\rangle \label{eq:GA} \\
& &\hspace{1cm} =\sum_r\langle\varphi|(\Gamma_{\epsilon}\Phi_r)*\psi\rangle\langle\Phi_r^c|\cI\rangle
-(-1)^{\mathrm{GSO}(\psi)}\sum_r\langle\varphi|\psi *(\Gamma_{\epsilon}\Phi_r)\rangle\langle\Phi_r^c|
\cI\rangle \nonumber \\
& &\hspace{1cm} =\sum_r(-1)^{\mathrm{GSO}(\psi)}\left\langle f_1^{(3)}\circ\psi(0) f^{(3)}_2\circ
\varphi(0)f^{(3)}_3\circ\Gamma_{\epsilon}\ f^{(3)}_3\circ\Phi_r(0)\right\rangle\langle f^{(1)}_1\circ
\Phi_r^c(0)\rangle \nonumber \\
& &\hspace{1.5cm}-(-1)^{\mathrm{GSO}(\psi)}\sum_r\left\langle f_1^{(3)}\circ\varphi(0) f^{(3)}_2\circ
\psi(0)f^{(3)}_3\circ\Gamma_{\epsilon}\ f^{(3)}_3\circ\Phi_r(0)\right\rangle\langle f^{(1)}_1\circ
\Phi_r^c(0)\rangle \nonumber \\
& &\hspace{1cm}=(-1)^{\mathrm{GSO}(\psi)}\left\langle F_1\circ f_1^{(3)}\circ\psi(0)F_1\circ f_2^{(3)}
\circ\varphi(0)F_1\circ f^{(3)}_3\circ\Gamma_{\epsilon}\right\rangle \nonumber \\
& &\hspace{1.5cm}-(-1)^{\mathrm{GSO}(\psi)}\left\langle F_1\circ f_1^{(3)}\circ\varphi(0)F_1\circ f_2^{(3)}
\circ\psi(0)F_1\circ f^{(3)}_3\circ\Gamma_{\epsilon}\right\rangle, \nonumber
\end{eqnarray}
where we have inserted the complete set of states, $\mathbf{1}=\sum_r|\Phi_r\rangle\langle\Phi_r^c|$, 
and we are omitting the symbol $\lim_{\epsilon\to 0}$. From the GGRT formulas given in~\cite{KisOhm} 
$F$'s are found to be 
\begin{eqnarray*}
F_1(z)&=&h^{-1}\left( e^{\frac{\pi}{2}i}h(z)^{\frac{3}{2}}\right), \\
\widehat{F}_2(z)&=&h^{-1}\left( e^{\frac{3\pi}{2}i}h(z)^{\frac{1}{2}}\right). 
\end{eqnarray*}
Calculating the compositions of the conformal maps carefully and using the $SL(2,\aaru)$-invariance of 
the correlators, we obtain 
\begin{eqnarray}
\langle\varphi |\Qev |\psi\rangle&=&(-1)^{\mathrm{GSO}(\psi)}\left\langle\varphi(0)\cR_{\pi}\circ
\Gamma_{\epsilon}\ \cR_{-\pi}\circ\psi(0)\right\rangle \nonumber \\
& &{}-(-1)^{\mathrm{GSO}(\psi)}\left\langle\varphi(0)\cR_{2\pi}\circ\Gamma_{\epsilon}\ \cR_{\pi}\circ
\psi(0)\right\rangle \nonumber \\
&=&\langle\varphi|\left(\Gamma_{\epsilon}-(-1)^{\mathrm{GSO}(\psi)}\cR_{\pi}\circ\Gamma_{\epsilon}\right)|\psi\rangle.
\phantom{QQQQQQQQQQQQQQ} \label{eq:GB}
\end{eqnarray}
Now we can take the $\epsilon\to 0$ limit without encountering any singularities. From the definition~(\ref{eq:BI}), 
(\ref{eq:BJ}) of the action of $\cR_{\pi}$ we find 
\begin{eqnarray}
& &\lim_{\epsilon\to 0}\left(\Gamma_{\epsilon}-(-1)^{\mathrm{GSO}(\psi)}\cR_{\pi}\circ\Gamma_{\epsilon}\right)
=q_1\frac{1-i}{4\varepsilon_r}\biggl\{\gamma(i)+i\gamma(-i) \nonumber \\
& &\hspace{5cm}-(-1)^{\mathrm{GSO}(\psi)}\left( i\gamma(\cR_{\pi}(i))+i(-i)\gamma(\cR_{\pi}(-i))\right)\biggr\}
\nonumber \\ & &\hspace{2cm}=q_1\frac{1-i}{4\varepsilon_r}(1-i(-1)^{\mathrm{GSO}(\psi)})\left(\gamma(i)-
(-1)^{\mathrm{GSO}(\psi)}\gamma(-i)\right), \label{eq:GC}
\end{eqnarray}
which completes the proof of (\ref{eq:CI}).

\sectiono{Twist Invariance of the Cubic Action}\label{sec:appB}
Here we explicitly show that the cubic action~(\ref{eq:CA}) with our choice of $\widehat{\cQ}$ 
is invariant under the twist transformation $\Omega$~(\ref{eq:AZ}). Since the cubic interaction term is not 
changed from the original one~(\ref{eq:AA}) at all, its twist invariance is shown in the same way as in~\cite{0011117}. 
Hence we will concentrate on the quadratic term. For $\Qod$ we consider 
\begin{eqnarray}
& &\llk Y_{-2}|\cO_1,\Qod\cO_2\rrk=\langle Y(i)Y(-i)\ (f^{(2)}_1\circ\cO_1(0))(f^{(2)}_2\circ\Qod)(f^{(2)}_2\circ\cO_2(0))
\rangle_{\mathrm{UHP}} \nonumber \\ & &\hspace{1.5cm}=\langle Y(0)Y(\infty)\ (g^{(2)}_1\circ\cO_1(0))(g^{(2)}_2\circ
\Qod)(g^{(2)}_2\circ\cO_2(0))\rangle_{\mathrm{disk}}, \label{eq:GD}
\end{eqnarray}
where $\langle\ldots\rangle_{\mathrm{disk}}$ denotes the correlation function on the unit disk, and 
\[ g^{(2)}_1(z)=h\circ f^{(2)}_1(z)=h(z),\quad g^{(2)}_2(z)=h\circ f^{(2)}_2(z)=e^{\pi i}h(z). \] 
Note that for the expression~(\ref{eq:GD}) 
to be non-vanishing $\cO_1$ and $\cO_2$ must live in the same GSO sector. Defining $M(z)\equiv e^{i\pi}z$, we find the 
following relations for any primary vertex operator $\varphi(z)$: 
\begin{eqnarray}
g^{(2)}_1\circ M\circ\varphi(z)&=&\tilde{I}\circ\cR_{2\pi}\circ g^{(2)}_1\circ\varphi(z), \nonumber \\
g^{(2)}_2\circ M\circ\varphi(z)&=&\tilde{I}\circ g^{(2)}_2\circ\varphi(z), \label{eq:GE}
\end{eqnarray}
where $\tilde{I}(z)\equiv 1/z,[\tilde{I}^{\prime}(z)]^h\equiv e^{-i\pi h}z^{-2h}$, and 
$\cR_{2\pi}(z)=e^{2\pi i}z$. If we assume that $\cO_1,\cO_2$ are primary fields of 
conformal weight $h_1,h_2$ respectively, we have 
\begin{eqnarray}
\cO_1(0)&=&e^{-i\pi h_1}M\circ\cO_1(0), \label{eq:GFa} \\ \cO_2(0)&=&e^{-i\pi h_2}M\circ\cO_2(0), \nonumber
\end{eqnarray}
\begin{eqnarray}
M\circ\Qod&=&\frac{1}{2i\varepsilon_r^2}(M\circ c(i)-M\circ c(-i))+\frac{q_1^2}{2}\oint\frac{dz}{2\pi i}
M\circ (b\gamma^2(z)) \nonumber \\
&=&\frac{1}{2i\varepsilon_r^2}\left( e^{-i\pi}c(ie^{i\pi})-e^{-i\pi}c(-ie^{i\pi})\right)
+\frac{q_1^2}{2}\oint\frac{dz}{2\pi i}e^{\pi i}b\gamma^2(e^{\pi i}z) \nonumber \\
&=&\frac{1}{2i\varepsilon_r^2}(c(i)-c(-i))+\frac{q_1^2}{2}\oint\frac{dz^{\prime}}{2\pi i}b\gamma^2
(z^{\prime})=\Qod. \label{eq:GF}
\end{eqnarray}
Then (\ref{eq:GD}) can be rewritten as 
\begin{eqnarray}
& &\llk Y_{-2}|\cO_1,\Qod \cO_2\rrk \nonumber \\ & &\hspace{1cm}
=e^{-i\pi(h_1+h_2)}\langle Y(0)Y(\infty)(g^{(2)}_1\circ M\circ\cO_1(0))
(g^{(2)}_2\circ M\circ\Qod) (g^{(2)}_2\circ M\circ\cO_2(0))\rangle_{\mathrm{disk}} \nonumber \\
& &\hspace{1cm}=e^{-i\pi(h_1+h_2)}\langle\tilde{I}\circ\left(Y(0)Y(\infty)\ (\cR_{2\pi}\circ g^{(2)}_1\circ \cO_1(0))
(g^{(2)}_2\circ\Qod) (g^{(2)}_2\circ\cO_2(0))\right)\rangle_{\mathrm{disk}} \nonumber \\
& &\hspace{1cm}=(-1)^{h_1+h_2}(-1)^{2h_1}\langle Y(0)Y(\infty)\ g^{(2)}_1\circ\cO_1(0)\ g^{(2)}_2
\circ (\Qod \cO_2(0))\rangle_{\mathrm{disk}} \nonumber \\
& &\hspace{1cm}=(-1)^{h_1+h_2}(-1)^{2h_1}\llk Y_{-2}|\cO_1,\Qod\cO_2\rrk, \label{eq:GG}
\end{eqnarray}
where we have used (\ref{eq:GE})--(\ref{eq:GF}), the invariance of $Y(0)Y(\infty)$ under the conformal transformation 
$\tilde{I}$, the invariance of the disk correlator under the $SL(2,\aaru)$ map 
$\tilde{I}(z)$, and $\cR_{2\pi}\circ\cO_1(0)=e^{2\pi ih_1}\cO_1(0)$. $(-1)^{h_1+h_2}$ and $(-1)^{2h_1}$ 
are well-defined because $h_1+h_2$ and $2h_1$ are always integers. Since eq.(\ref{eq:GG}) says that 
$\llk Y_{-2}|\cO_1,\Qod\cO_2\rrk$ is equal to itself multiplied by $(-1)^{h_1+h_2}(-1)^{2h_1}$, this expression 
vanishes unless $(-1)^{h_1+h_2}(-1)^{2h_1}=1$ holds. This condition is satisfied only when both $\cO_1$ and 
$\cO_2$ are twist-odd or twist-even. For example, consider a case $h_1\in 2\zetto +\frac{1}{2}$ (\textit{i.e.} 
$\cO_1$ is \textit{twist-odd} and in the GSO($-$) sector). Since $(-1)^{2h_1}=-1$, $h_2$ must take value in 
$2\zetto +\frac{1}{2}$ so that $(-1)^{h_1+h_2}$ will equal $-1$. In other words, $\cO_2$ must also be 
\textit{twist-odd} and in the GSO($-$) sector. One can repeat similar arguments for other three cases to 
verify the above statement. To sum up, we have shown that twist-odd states enter $\llk Y_{-2}|\cO_1,\Qod\cO_2\rrk$ 
always \textit{in pairs} (otherwise the expression vanishes), so that it is invariant under the twist operation 
$\Omega$. 

Now we turn to 
\begin{equation}
\llk Y_{-2}|\cO_1^{(+)},\Qev^{\mathrm{GSO}(-)}\cO_2^{(-)}\rrk=\langle Y(0)Y(\infty)\ g^{(2)}_1\circ\cO_1^{(+)}(0)\ 
g^{(2)}_2\circ(\Qev^{\mathrm{GSO}(-)}\ \cO_2^{(-)}(0))\rangle_{\mathrm{disk}}, \label{eq:GH}
\end{equation}
where the superscripts $(\pm)$ for $\cO_1^{(+)},\cO_2^{(-)}$ denote their GSO parities. In a similar way to the 
above argument, we rewrite~(\ref{eq:GH}) as 
\begin{eqnarray}
\llk Y_{-2}|\cO_1^{(+)},\Qev^{\mathrm{GSO}(-)}\cO_2^{(-)}\rrk&=&e^{-i\pi(h_1+h_2-\frac{1}{2})}\langle Y(0)Y(\infty)\ 
(g^{(2)}_1\circ M\circ\cO_1^{(+)}(0)) \nonumber \\ & &{\times}
(g^{(2)}_2\circ M\circ\Qev^{\mathrm{GSO}(-)}) (g^{(2)}_2\circ M\circ\cO_2^{(-)}(0))\rangle_{\mathrm{disk}} \nonumber \\
&=&(-1)^{h_1+h_2-\frac{1}{2}}(-1)^{2h_1}\llk Y_{-2}|\cO_1^{(+)},\Qev^{\mathrm{GSO}(-)}\cO_2^{(-)}\rrk,\phantom{QQQQQ} 
\label{eq:GI}
\end{eqnarray}
where we have used 
\[ M\circ\Qev^{\mathrm{GSO}(-)}=\frac{q_1}{2\varepsilon_r}(e^{i\pi})^{-\frac{1}{2}}(\gamma(ie^{i\pi})+
\gamma(-ie^{i\pi}))=e^{-\frac{\pi i}{2}}\Qev^{\mathrm{GSO}(-)}. \]
Since $\cO_1^{(+)}$ lives in the GSO($+$) sector, $(-1)^{2h_1}=1$. If $\cO_1^{(+)}$ is twist-odd, \textit{i.e.} 
$h_1\in 2\zetto$, then $(-1)^{h_1+h_2-\frac{1}{2}}=1$ requires $h_2\in 2\zetto +\frac{1}{2}$ which means that 
the GSO($-$) state $\cO_2^{(-)}$ should be twist-odd. Conversely, if $\cO_1^{(+)}$ is twist-even, $\cO_2^{(-)}$ 
must be twist-even for $(-1)^{h_1+h_2-\frac{1}{2}}=1$ to hold. 
We again conclude that twist-odd states appear in non-zero 
$\llk Y_{-2}|\cO_1^{(+)}, \Qev^{\mathrm{GSO}(-)}\cO_2^{(-)}\rrk$ \textit{pairwise}  so that 
this expression is twist invariant. Finally we consider 
\begin{eqnarray}
& &\llk Y_{-2}|\cO_1^{(-)},\Qev^{\mathrm{GSO}(+)}\cO_2^{(+)}\rrk=\langle Y(0)Y(\infty)\ g^{(2)}_1\circ\cO_1^{(-)}(0)\ 
g^{(2)}_2\circ(\Qev^{\mathrm{GSO}(+)}\ \cO_2^{(+)}(0))\rangle_{\mathrm{disk}} \nonumber \\
& &\hspace{1cm}=e^{-i\pi (h_1+h_2+\frac{1}{2})}\langle Y(0)Y(\infty)\ (g^{(2)}_1\circ M\circ\cO_1^{(-)}(0)) \nonumber \\
& &\hspace{2cm}\times 
(g^{(2)}_2\circ M\circ\Qev^{\mathrm{GSO}(+)})(g^{(2)}_2\circ M\circ\cO_2^{(+)}(0))\rangle_{\mathrm{disk}} \nonumber \\
& &\hspace{1cm}=(-1)^{h_1+h_2+\frac{1}{2}}(-1)^{2h_1}\llk Y_{-2}|\cO_1^{(-)},\Qev^{\mathrm{GSO}(+)}
\cO_2^{(+)}\rrk, \label{eq:GJ}
\end{eqnarray}
where we have used 
\[ M\circ\Qev^{\mathrm{GSO}(+)}=\frac{q_1}{2i\varepsilon_r}(e^{i\pi})^{-\frac{1}{2}}(\gamma(-i)-\gamma(i))
=-e^{-\frac{i\pi}{2}}\Qev^{\mathrm{GSO}(+)}\equiv e^{\frac{i\pi}{2}}\Qev^{\mathrm{GSO}(+)}. \]
The argument goes along the same line: If $h_1\in 2\zetto+\frac{1}{2}$ ($\cO_1^{(-)}$ is twist-odd) then we 
must have $h_2\in 2\zetto$ ($\cO_2^{(+)}$ is twist-odd) because of $(-1)^{2h_1}=-1$. Putting the above 
considerations together, we have completed the proof that the quadratic part 
$\mathrm{Tr}\llk\widehat{Y}_{-2}|\widehat{\cA},\widehat{\cQ}\widehat{\cA}\rrk$ of the action with 
our $\widehat{\cQ}$ is twist invariant.


\end{document}